\documentclass[useAMS,usenatbib]{mn2e}
\usepackage{amssymb}
\usepackage{graphicx}
\usepackage{mathrsfs}
\usepackage{xcolor}
\newcommand{\chem}[2]{$\mathrm{^{#1}#2}$}
\def\Msun{$M_\odot$\,}
\def\msun{$M_\odot$}

\def\rsun{$R_\odot$}

\title[Long-rising SNe: a study through scaling relations]{Long-rising Type II supernovae resembling supernova 1987A – I. A comparative study through scaling relations}

\author[M.~L.~Pumo et al.]{M.~L.~Pumo$^{1,2,3}$\thanks{E-mail: marialetizia.pumo@unict.it}, S.~P.~Cosentino$^{1}$, A.~Pastorello$^{2}$, S.~Benetti$^{2}$, S.~Cherubini$^{1,3}$,\\
\newauthor 
G.~Manic\`o$^{1,3}$, L.~Zampieri$^{2}$\\
$^{1}$Universit\`a degli studi di Catania, Dip.~di Fisica e Astronomia ``Ettore Majorana'', Catania, Italy\\ 
$^{2}$INAF - Osservatorio Astronomico di Padova, Padova, Italy\\
$^{3}$Laboratori Nazionali del Sud-INFN, Catania, Italy} 
\begin{document}

\date{Accepted 2023 March 14. Received 2023 February 25; in original form 2022 August 1.}

\pagerange{\pageref{firstpage}--\pageref{lastpage}} \pubyear{....}

\maketitle

\label{firstpage}

\begin{abstract}
With the aim of improving our knowledge about their nature, we conduct a comparative study on a sample of long-rising Type II supernovae (SNe) resembling SN 1987A. To do so, we deduce various scaling relations from different analytic models of H-rich SNe, discussing their robustness and feasibility. Then we use the best relations in terms of accuracy to infer the SN progenitor's physical properties at the explosion for the selected sample of SN 1987A-like objects, deriving energies of $\sim 0.5$-$15$ foe, radii of $\sim 0.2$-$100 \times 10^{12}$ cm, and ejected masses of $\sim 15$-$55$\msun. Although the sample may be too small to draw any final conclusion, these results suggest that (a) SN 1987A-like objects have parameters at explosion covering a wide range of values; (b) the main parameter determining their distribution is the explosion energy; (c) a high-mass ($\gtrsim 30$\,\Msun), high-energy ($\gtrsim 10$\,foe) tail of events, linked to extended progenitors with radii at explosion $\sim 10^{13}$-$10^{14}$\,cm, challenge standard theories of neutrino-driven core-collapse and stellar evolution. We also find a correlation between the amount of $^{56}$Ni in the ejecta of the SN 1987A-like objects and the spectrophotometric features of the SN at maximum, that may represent a tool for estimating the amount of $^{56}$Ni in the SN ejecta whitout having information on the tail luminosity.
\end{abstract}                                                                                                                                                                                                                                                                                                                                                                                                                                                                                                                                                                                                                                                                                                                                                                                                                                                                                                                                                                                                                                   


\section{Introduction}
\label{intro}

It is widely accepted that supernova (SN) 1987A-like objects form a subclass of Type II SNe characterized by long-rising (exceeding 40-50 days) bolometric light curves with shapes resembling that of SN 1987A \citep[e.g.][and references therein]{taddia16}. These explosive events seem to be intrinsically rare \citep[$\lesssim$ 1-3 per cent of all core-collapse SNe in a volume-limited sample; e.g.][]{smartt09,kleiser11,pasto12,taddia16} and, at present, a few tens of objects have been classified as belonging to this SN sub-group \citep[e.g.][and references therein]{takats16}.\par

The long-rising SNe usually show bolometric luminosities at the peak ranging from $\sim 3$-$5\times10^{41}$ to $\sim 3$-$5\times10^{42}$\,erg\,s$^{-1}$, \chem{56}{Ni} masses powering their tail luminosity in the range $\sim 0.05$-$0.25$\,\msun, and spectra with P-Cygni lines similar to those of ``normal'' Type II SNe \citep[e.g.][]{pasto05,kleiser11,taddia12,pasto12,taddia16,takats16}.\par

All these features are usually explained in terms of core-collapse explosions with energies in the range $\sim 0.5$-$5$\,foe (1 foe $\equiv 10^{51}$\,ergs), occurring in relatively compact (radius at explosion $\sim 30$-$300$\,\rsun) and massive (ejected mass $\sim 12$-$30$\,\Msun) progenitors \citep*[e.g.][]{woosley88,arnett89,SN90,utrobin95,blinnikov00,PZ11,uc11,taddia12,pasto12,PZ13,orlando15,taddia16,takats16}. However \citet[][]{taddia16} suggest that progenitors with very extended radii (of the order of thousands of \rsun) could also produce long-rising SNe if a sufficiently large amount of \chem{56}{Ni} ($\gtrsim 0.1$-$0.2$\,\Msun) is synthesized during the SN explosion.\par

In this context, the discovery of OGLE-2014-SN-073 (hereafter referred to as OGLE073) during the OGLE-IV\footnote{http://ogle.astrouw.edu.pl/ogle4/transients/} survey \citep[][]{wyrzykowski14} is of remarkable importance. Indeed, in addition to belonging to the rare group of SN 1987A-like objects, OGLE073 shows very peculiar features \citep[see][for details]{terreran16}: (a) it is the brightest SN 1987A-like object ever discovered ($\sim$ 3 mag more luminous than the prototype of the class SN 1987A and the second brightest non-interacting Type II SN after SN 2009kf); (b) its \chem{56}{Ni} mass of at least $\sim$ 0.45\,\Msun\,is the largest ever estimated for a long-rising SN and, more in general, for a Type II SN; and (c) analyses based on radiation-hydrodynamical models of SN ejecta, indicate that the physical properties of its progenitor at explosion (primary the ejected mass and the explosion energy) are difficult to explain within the conventional neutrino-driven core-collapse paradigm. These results, together with those obtained by \citet[][]{taddia16} for SNe 2004ek and 2004em, seem to indicate the existence of Ni-rich ($\gtrsim 0.1$-$0.2$\,\Msun), high-mass ($\gtrsim 30$\,\Msun), high-energy ($\gtrsim 10$\,foe) events forming a luminous tail of SN 1987A-like objects, that could be characterised by a ``non-conventional'' explosion. On the other hand, the existence of faint clones of SN 1987A as SN 2009E \citep[][]{pasto12} or the more ``extreme'' and enigmatic objects such as SN DES16C3cje \citep[][]{gutierrez20}, show the possible presence of a sub-luminous tail of SN 1987A-like objects.\par

Slowly rising SNe seem thus to form a group of objects with a distribution in the parameter space analogously to that found for Type II plateau SNe \citep[e.g.][]{zampieri07,spiro14,anderson14,faran14,sanders15,pumo17}, as already suggested by \citet[][]{pasto12}. However a comparative study among long-rising SNe focused on verifying the possible existence of systematic trends inside this sub-group of SNe, is still missing.\par

With the aim of studying systematics within the SN 1987-like objects's family, based on analytic models describing the post-explosive evolution of SN ejecta for H-rich events, we derive different scaling relations that enable us to infer the SN progenitor's physical properties at the explosion (namely the ejected mass $M_{ej}$, the progenitor radius at the explosion $R$ and the total explosion energy $E$) for long-rising SNe. After testing the robustness of these relations (most of which are new), we apply the best ones in terms of accuracy and precision to one of the biggest and most complete sample of well-observed SN 1987A-like objects ever considered in the literature. A preliminary analysis of this type was carried out by \citet[][]{taddia16} using a less refined approach on a more limited sample of SN 1987A-like objects.\par
  
The plan of the paper is the following. We illustrate the sets of scaling relations in Section \ref{modelling} and briefly present the models used for testing purpose as well as the sample of SN 1987A-like objects in Section \ref{samples}. In Section \ref{results} we present and discuss our results, devoting Section \ref{scaling_rel} to the scaling relations and Section \ref{systematics} to the comparative study. A summary is presented in Section \ref{summary}.\par

\section{Sets of scaling relations}
\label{modelling}
A first set of scaling relations can be obtained using the simple ``one zone'' model of \citet[][]{arnett80}. In addition to the spherical symmetry, this model hypothesizes ejecta of uniform density in homologous expansion and having a radiation-dominated energy density (with initial energy nearly equally divided between kinetic and thermal). Further assumptions of the model are: (i) radiative diffusion, (ii) constant opacity, and (iii) neglecting both the effects of recombination and the heating due to the decay of radioactive isotopes synthesized during the explosion. Under these assumptions, the following relations hold:
\begin{equation}
\label{eq1}
E_{th,0} \simeq E/2\propto M_{ej}\>v_{sc}^2,
\end{equation}
\begin{equation}
\label{eq2}
t_a=\sqrt{2\>t_e\>t_d}\propto (\kappa\>M_{ej}/v_{sc})^{1/2},
\end{equation}
\begin{equation}
\label{eq3}
L(t)\simeq \frac{E_{th,0}}{t_d} \cdot \exp(-t^2/t_a^2) \simeq \frac{E}{2t_d} \cdot \exp(-t^2/t_a^2),
\end{equation}
where $E_{th,0}$ is the thermal energy produced during the collapse, $v_{sc}$ is the so-called velocity scale which corresponds to the velocity of the ejecta's outer layer\footnote{In literature the scale velocity is frequently used to describe the expansion velocity of the SN ejecta \citep*[e.g.][]{arnett80,chugai91,popov93,balberg00,KW09,chatzopoulos12,KK19}. In particular, given the omologous explosion, the velocity of a Lagrangian particle of the ejecta at a distance $r$ from the centre and at the time after the explosion $t$ is $$v(r,t)= x v_{sc},$$ where $x$ is the dimensionless radius \citep[see also Eq.~28 of][]{arnett80}.}, $\kappa$ is the opacity, $L(t)$ is the bolometric luminosity at the generic time from the explosion $t$, $t_a$ is the timescale necessary for cooling down the structure by a factor of $e$, and $t_e= R/v_{sc}$ and $t_d \propto kM_{ej}/R$ are other two characteristic timescales, usually labeled as expansion time and diffusion time, respectively.\par 

Adopting a similar opacity $\kappa$ for all long-rising SNe, the time (from the explosion) needed to reach the bolometric peak $t_M$\footnote{As a general rule throughout the manuscript, unless differently specified, the variables with a capital letter ``M'' as subscript refer to quantities estimated from observational data and evaluated when the bolometric light curve is at maximum. Similarly, variables with a lower case letter ``m'' as subscript refer to quantities estimated from observational data and evaluated when the bolometric light curve is at a minimum.\label{footnote:letter}} as an estimate of $t_a$, and the photopheric velocity at peak $v_{ph}(t_M)\equiv v_M$ as a measure of $v_{sc}$, Eq.s (\ref{eq1})-(\ref{eq3}) can be rewritten to form the following set of scaling relations:
\begin{equation}
\label{setArnett}
\left\{
\begin{array}{l}
\displaystyle
E      \propto t_M^{2} v_M^3\\
M_{ej} \propto t_M^{2} v_M\\
R      \propto L_M v_M^{-2} \,\,\,\,\,\mbox{with}\,\,\, L_M \equiv L(t_M).
\end{array}
\right.
\end{equation}
\par
This set of proportional relationships links the values of $E$, $M_{ej}$, and $R$ to a series of parameters (namely, $t_M$, $v_M$, and $L_M$) that depend on the spectro-photometric behavior of the SN at the epoch of the bolometric light-curve maximum. Therefore, once such behavior is known for a sample of long-rising SNe, it is possible to obtain $E$, $M_{ej}$, and $R$ in a homogeneous way for all the sample, provided that $E$, $M_{ej}$, and $R$ can be independently evaluated at least for one SN of the sample (hereafter referred to as reference SN). Note that the first two relations of set (\ref{setArnett}) are also valid for radioactive SNe \citep[][]{arnett79} and have been sometimes used to estimate the values of $E$ and $M_{ej}$ for various SN 1987A-like events \citep[e.g.][]{kleiser11,taddia12,taddia16}.\par

Other new sets of scaling relations can be obtained using the ``two zone'' model of \citet[][]{popov93}. This model is essentially based on the same assumptions used in \citet[][]{arnett80}, but the effects of recombination are taken into account. In particular, it is assumed that the recombination of the ejected material occurs at ionization temperature $T_i$, and the opacity is approximated with the following staircase function of the temperature:

\begin{equation}
\label{opacity}
\kappa=  
\left\{
\begin{array}{l}
\displaystyle
\mbox{constant} \neq 0 \,\,\,\,\,\mbox{if}\,\, T \geq T_i\\
\\
0 \,\,\,\,\,\,\,\,\,\,\,\,\,\,\,\,\,\,\,\,\,\,\,\,\,\,\,\,\,\,\,\,\,\,\,\, \mbox{if}\,\, T < T_i.
\end{array}
\right.
\end{equation}
\noindent In this way, a recombination front moving inward (in mass), marks the photosphere and divides the ejecta into two regions: an inner part that is optically thick, ionized and hot ($T \geq T_i$), and an outer zone that is optically thin, recombined and cooler. Under these assumptions, relations (\ref{eq1}) and (\ref{eq2}) remain valid, but the photopheric velocity $v_{ph}$ is no longer a proxy value for $v_{sc}$, given the following relation \citep[see also Eq.s 4 and 15 of][]{popov93}:
\begin{equation}
\label{eqVsc}
v_{ph}= v(x_i)= x_i(t) v_{sc}= \bigg[\frac{t_i}{t}\>\bigg(1+\frac{t_i^2}{3t_a^2}\bigg) -\frac{t^2}{3t_a^2}\bigg]^{1/2} v_{sc}
\end{equation}
where $x_i(t)$ is the dimensionless radius of the recombination front at the generic time from the explosion $t$, and $t_i$ is the time when the surface temperature decreases to $T_i$. As for the bolometric luminosity, equation (\ref{eq3}) remains valid prior to recombination (i.e. for $t < t_i$), thereafter the bolometric luminosity is given by the following relation \citep[see also Eq.~17 of][]{popov93}:
\begin{equation}
\label{eqLbolpopov}
L(t)= 8\pi \sigma_{SB}\>T^4_{i}\>v_{sc}^2\>\bigg[t_i\>t\>\bigg(1+\frac{t_i^2}{3t_a^2}\bigg)-\frac{t^4}{3t_a^2}\bigg]
\end{equation}
\noindent where $\sigma_{SB}$ is the Stefan-Boltzmann constant, and the maximum of the function $L(t)$ occurs at
\begin{equation}
\label{eqt_Lbolmax}
t_{max}= \bigg[\frac{3}{4} t_i t_a^2  \bigg(1 + \frac{t_i^2}{3t_a^2}\bigg) \bigg]^{1/3}  
\end{equation}
\noindent \citep[see also Eq.~18 of][]{popov93}. Matching equation (\ref{eq3}) with equation (\ref{eqLbolpopov}) for $t= t_i$, and considering the typical values of parameters describing the SN progenitor's physical properties appropriate to H-rich SNe (i.e.~Type II plateau SNe and SN 1987A-like objects), one obtains
\begin{equation}
\label{eqt_i}
t_i \propto \frac{R^{1/2}}{\kappa^{1/2} T_i^{2}}
\end{equation}
\noindent \citep[see also Eq.~25 of][]{popov93}. Moreover, considering once again the typical values of parameters describing the SN progenitor's physical properties appropriate to H-rich SNe, $t_i$/$t_a$ can be written as 
\begin{equation}
\label{t_i_over_t_a}
\frac{t_i}{t_a} \propto \frac{E^{1/4} R^{1/2}}{M_{ej}^{3/4}}
\end{equation}
\noindent \citep[see also Eq.s 24 and 25 of][]{popov93}, and the term $t_i^2$/$3t_a^2$ in equations (\ref{eqVsc}), (\ref{eqLbolpopov}), and (\ref{eqt_Lbolmax}) can be neglected because it is $\ll 1$.\par

Adopting similar values of $\kappa$ and $T_i$ for all long-rising SNe, and considering respectively the values of $t_M$, $L_M$, and $v_M$ as a measure of $t_{max}$, $L(t_{max})$, and $v_{ph}(t_{max})$, Eq.s (\ref{eqVsc})-(\ref{eqt_Lbolmax}) can be rewritten --- using also relations (\ref{eq1}), (\ref{eq2}), (\ref{eqt_i}), and (\ref{t_i_over_t_a}) --- to provide the following set of relations:
\begin{equation}
\label{setPpovdegenerate}
\left\{
\begin{array}{l}
\displaystyle
(a)\> \> E^3 M_{ej}^{-5} \propto v_M^4 t_M^{-4}\\
(b)\> \> R^2 M_{ej} E    \propto L_M^{3} v_M^{-2} t_M^2\\
(c)\> \> R^3 M_{ej}^4    \propto v_M^4 t_M^{14}. 
\end{array}
\right.
\end{equation}
\noindent In contrast with set (\ref{setArnett}), this is degenerate because it is equivalent to a 3x3 linear system with the determinant of the coefficient matrix equal to zero, and where the first equation of the system [corresponding to the relation (a) in set (\ref{setPpovdegenerate})] is a combination of the remaining ones [corresponding to the relations (b) and (c) in set (\ref{setPpovdegenerate})]. In particular, the following relation is valid (see Appendix \ref{app_formulas} for further details):
\begin{equation}
\label{eqForNi}
(a) \propto (b)^3 (c)^{-2}.
\end{equation}
As a consequence, set (\ref{setPpovdegenerate}) cannot be used to derive the values of $E$, $M_{ej}$, and $R$. In order to remove the degeneration and, thus, to estimate $E$, $M_{ej}$, and $R$, it is necessary to replace one of the three relations of set (\ref{setPpovdegenerate}) with another independent relation. Such relation should link the SN progenitor's physical properties at the explosion to physical quantities that do not solely depend on the SN spectro-photometric behavior at maximum. In particular, instead of the relation (c) of set (\ref{setPpovdegenerate}) derived from the estimate of $L$ at $t=t_{max}$, we use the corresponding relation that can be inferred from the estimate of $L$ at $t=t_i$. Thus, adopting the epoch of the bolometric light-curve minimum $t_m$ occurring prior to the rising stage as an estimate of $t_i$ and, consequently, using $L_m\equiv L(t_m)$ as a measure of $L_{i}\equiv L(t_i)$, one obtains
\begin{equation}
\label{eqLbolminpopov}
L_m= 8\pi \sigma_{SB} T^4_{i} v_{sc}^2 t_i^2 \propto v_M^{4/3} t_M^{-4/3} M^{2/3} R
\end{equation}
\noindent or
\begin{equation}
\label{eqLbolminpopovbis}
M^2 R^3 \propto L_m^3 v_M^{-4} t_M^4.
\end{equation}
This relation coupled with relations (b) and (c) of set (\ref{setPpovdegenerate}), constitute a not-degenerate system, that can be rewritten to form the following new set of scaling relations:
\begin{equation}
\label{setPopov}
\left\{
\begin{array}{l}
\displaystyle
E      \propto L_m^{-5/2} t_M^{7}  v_M^8\\
M_{ej} \propto L_m^{-3/2} t_M^{5}  v_M^4\\
R      \propto L_m^{2}    t_M^{-2} v_M^{-4}
\end{array}
\right.
\end{equation}
\noindent or, using relation (\ref{eqForNi}) (which implies that $v_M \propto L_M^{1/2} t_M^{-1}$; see also Appendix \ref{app_formulas} for further details),
\begin{equation}
\label{setPopov_phot}
\left\{
\begin{array}{l}
\displaystyle
E      \propto L_m^{-5/2} L_M^4    t_M^{-1}\\
M_{ej} \propto L_m^{-3/2} L_M^2    t_M     \\
R      \propto L_m^{2}    L_M^{-2} t_M^2.
\end{array}
\right.
\end{equation}
\par As set (\ref{setArnett}), sets (\ref{setPopov}) and (\ref{setPopov_phot}) enable us to derive $E$, $M_{ej}$, and $R$ from three parameters ($L_m$, $t_M$, and $v_M$ or $L_m$, $L_M$, and $t_M$) linked to the spectro-photometric behavior of long-rising SN. However, before using either the relations of set (\ref{setArnett}) or those of sets (\ref{setPopov}) and (\ref{setPopov_phot}), is necessary to carefully weigh pros and cons of each set. For example, set (\ref{setArnett}) has the clear advantage that it can be used once known the spectro-photometric behavior of the long-rising SN only\footnote{Of course it is also needed a sufficiently precise estimate of the phase since explosion for determining $t_M$.} at the epoch of the bolometric light-curve maximum, but the derived value of $R$ should be considered a rough estimate of its real measure \citep[see e.g.][and also Section \ref{scaling_rel}]{kleiser11}. In contrast, when $R$ is estimated using the corresponding relation of set (\ref{setPopov}), its value is likely more accurate, but it is necessary to know $L_m$ and, in turn, to well sample the bolometric light-curve also long before the rise to the maximum. Set (\ref{setPopov_phot}) has instead the clear advantage that it can be used when only the photometric behavior of the long-rising SN is known. However it is still necessary to well sample the pre-maximum light-curve, and the derived measures of $E$, $M_{ej}$, and $R$ should be considered a rough estimate of their real values (see also Section \ref{scaling_rel}).\par

In order to analyse the robustness of the scaling relations of sets (\ref{setArnett}), (\ref{setPopov}) and (\ref{setPopov_phot}) and, consequently, to determine the best ones in terms of accuracy and precision, we check them against a group of well-observed SN 1987A-like objects for some of which the values of $E$, $M_{ej}$, and $R$ were already inferred through the hydrodynamical modelling of the main SN observables (i.e. bolometric light curve, evolution of line velocities and continuum temperature at the photosphere). Moreover, we use a grid of radiation-hydrodynamical models of long-rising SNe to evaluate the impact of neglecting the heating effects due to the nickel decay on the robustness and the feasibility of the scaling relations. We also use this grid of models to examine the effects of the choice of the reference SN on the estimation of $E$, $M_{ej}$, and $R$ for all sets of scaling relations. The grid of radiation-hydrodynamical models and the sample of well-observed SN 1987A-like objects are described in detail in Section \ref{samples}.
\par

\section{Samples of long-rising SNe}
\label{samples}

\subsection{Radiation-hydrodynamical models}
\label{models}
We consider a homogeneous grid of hydrodynamical computations, obtained using the general-relativistic, radiation-hydrodynamics Lagrangian code presented in \citet*[][]{pumo10} and \citet[][]{PZ11}. This code is specifically designed to simulate the evolution of the physical properties of the ejected material and the behavior of the main SN observables in core-collapse events. In particular it is able to follow the entire post-explosive evolution (i.e.~from the shock wave's breakout at the stellar surface up to the radioactive-decay phase), taking into account both the gravitational effects of the compact remnant and the heating effects due to the decays of the radioactive isotopes synthesized during the explosion. The basic parameters driving the post-explosion evolution of the modes are $M_{ej}$, $E$, $R$, and the $^{56}$Ni mass, $M_{Ni}$, initially present in the ejected material \citep[see also][]{PZ13}.\par
\begin{table}
   \centering
   \caption{Basic parameters of the radiation-hydrodynamical models (see text for details). Masses are in solar units, progenitor radius in $10^{12}$\,cm, and energy in foe ($\equiv$10$^{51}$\,ergs).}
   \begin{tabular}{ccccc}
   \hline\hline
   Model & $M_{ej}$ & $E$   & $R$           & $M_{Ni}$ \\ 
         & [\msun]  & [foe] & [$10^{12}$\,cm] & [\msun]\\    
   \hline
   $1$   & $16$     & $1$   & $3$           & $0.0001 $\\
   $2$   & $16$     & $1$   & $3$           & $0.001  $\\
   $3$   & $16$     & $1$   & $3$           & $0.01   $\\
   $4$   & $16$     & $1$   & $3$           & $0.04   $\\
   $5$   & $16$     & $1$   & $3$           & $0.07   $\\ 
   $6$   & $16$     & $1$   & $3$           & $0.1    $\\
   $7$   & $16$     & $1$   & $3$           & $0.25   $\\
   $8$   & $16$     & $1$   & $3$           & $0.5    $\\
  \hline
 \end{tabular}
 \label{tabmodels}
\end{table}
The grid is composted of 8 SN 1987A-like models, having the same parameters $M_{ej}$, $E$, and $R$, but different $M_{Ni}$. In particular, they have $M_{ej}= 16$\msun, $E= 1foe$, $R= 3$x$10^{12}cm$, and $M_{Ni}$ ranging from $10^{-4}$\Msun to $0.5$\Msun (see also Table \ref{tabmodels}).\par 

Using this grid, it is thus possible to get information about the impact of neglecting the heating effects due to the $^{56}$Ni decay on the accuracy of the relations of sets (\ref{setArnett}), (\ref{setPopov}) and (\ref{setPopov_phot}). This type of analysis cannot be appropriately done using real SN data neither considering models with different values of $M_{ej}$, $E$, and $R$ (in addition to different values of $M_{Ni}$) because, in these cases, it would be impossible to unequivocally constrain the $^{56}$Ni decay effects.\par

\subsection{Well-observed SN 1987A-like objects}
\label{observations}
The sample of well-observed SN 1987A-like objects is composed by the following 14 SNe: 1987A, 1998A, 2000cb, 2004ek, 2004em, 2005ci, 2006V, 2006au, 2009E, 2009mw, PTF12gcx, PTF12kso, OGLE073, and DES16C3cje. All the observational data used in the present work are taken from \citet[][and references therein]{taddia16}, except SNe 2009mw, OGLE073, and DES16C3cje whose observational data are taken from \citet[][]{takats16}, \citet[][]{terreran16}, and \citet[][]{gutierrez20}, respectively.\par

\begin{table*}
   \centering
   \caption{Selected parameters (see text for details) for our sample of SN 1987A-like objects. Luminosities are in $10^{40}$~erg~s$^{-1}$, velocity in km~s$^{-1}$, and time in days. Estimated uncertainties on the inferred parameters are in brackets. The Table is ordered following the brightness of the SNe at the epoch of the bolometric light-curve maximum, with the most luminous one being at the top.}
   \begin{tabular}{lllll}
   \hline\hline
   SN & \multicolumn{1}{c}{$L_m$} & \multicolumn{1}{c}{$L_M$} & \multicolumn{1}{c}{$v_M$} & \multicolumn{1}{c}{$t_M$}\\
      & \multicolumn{1}{c}{[$10^{40}$~erg s$^{-1}$]} & \multicolumn{1}{c}{[$10^{40}$~erg s$^{-1}$]} & \multicolumn{1}{c}{[km s$^{-1}$]} & \multicolumn{1}{c}{[d]}\\    
   \hline                                        
   OGLE073     & $ N.A.              $  & $ 1018 \,(\pm 100) $  & $5155 \,(\pm 358)  $    & $102 \,(\pm 9)  $\\
   2004ek     & $ 425  \,(\pm 25)   $  & $ 539  \,(\pm 8)   $  & $4834 \,(\pm 467)  $    & $80  \,(\pm 6)  $\\
   PTF12kso   & $ N.A.              $  & $ 320  \,(\pm 9)   $  & $4196 \,(\pm 1090) $    & $65  \,(\pm 1)  $\\
   PTF12gcx   & $ 70   \,(\pm 10)   $  & $ 225  \,(\pm 50)  $  & $4256 \,(\pm 1000) $    & $60  \,(\pm 25) $\\
   2004em     & $ 110  \,(\pm 15)   $  & $ 197  \,(\pm 61)  $  & $4122 \,(\pm 1217) $    & $89  \,(\pm 36) $\\ 
   2006V      & $ N.A.              $  & $ 179  \,(\pm 50)  $  & $2860 \,(\pm 404)  $    & $78  \,(\pm 7)  $\\
   2006au     & $ 91   \,(\pm 10)   $  & $ 157  \,(\pm 30)  $  & $4366 \,(\pm 1000) $    & $74  \,(\pm 3)  $\\
   1998A      & $ N.A.              $  & $ 113  \,(\pm 30)  $  & $2857 \,(\pm 70)   $    & $95  \,(\pm 5)  $\\
   1987A      & $ 16   \,(\pm 5)    $  & $ 96   \,(\pm 10)  $  & $2069 \,(\pm 93)   $    & $89  \,(\pm 5)  $\\
   2000cb     & $ \leq 15 \,(\pm 5) $  & $ 84   \,(\pm 26)  $  & $4231 \,(\pm 240)  $    & $68  \,(\pm 2)  $\\
   2005ci     & $ \leq 13 \,(\pm 5) $  & $ 68   \,(\pm 15)  $  & $2733 \,(\pm 79)   $    & $93  \,(\pm 2)  $\\
   2009mw     & $ \leq 21 \,(\pm 5) $  & $ 68   \,(\pm 3)   $  & $3004 \,(\pm 302)  $    & $86  \,(\pm 5)  $\\
   2009E      & $ N.A.              $  & $ 64   \,(\pm 15)  $  & $1539 \,(\pm 156)  $    & $95  \,(\pm 4)  $\\
   DES16C3cje & $ 14   \,(\pm 5)    $  & $ 46   \,(\pm 18)  $  & $1150 \,(\pm 900)  $    & $143 \,(\pm 12) $\\
  \hline                                   
 \end{tabular}
 \label{tab_observations}

 In the second column, $N.A.$ stands for ``not available'' because it is not possible to infer $L_m$ from the available observational data, and the symbol $\leq$ indicates that it is just an upper limit. 
 \end{table*}

\begin{figure}
 \includegraphics[angle=-90,width=85mm]{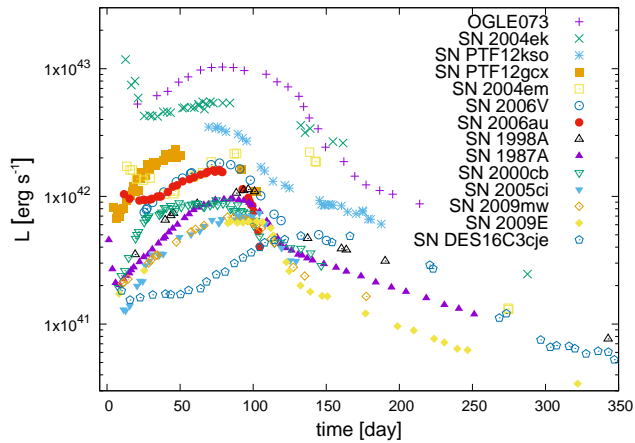} 
 \caption{(Pseudo-)Bolometric luminosities during the first 350 days after the explosion for our sample of SN 1987A-like objects.
 \label{figBol}}
\end{figure}

\begin{figure}
 \includegraphics[angle=-90,width=85mm]{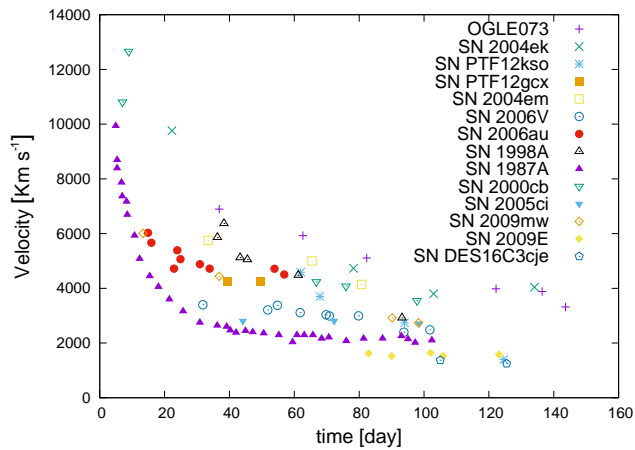} 
 \caption{Same as Figure \ref{figBol}, but for the behavior of the photospheric velocity.
 \label{figVel}}
\end{figure}

For this sample of SN 1987A-like objects, Table \ref{tab_observations} shows the values of the parameters $L_m$, $L_M$, $v_M$, and $t_M$ to be used in the relations of sets (\ref{setArnett}), (\ref{setPopov}) and (\ref{setPopov_phot}). These parameters are derived interpolating the data shown in Figures \ref{figBol} and \ref{figVel}, where the bolometric light curve and the photospheric velocity are respectively reported as a function of the phase for the considered sample. For the photospheric velocity, we use the values derived from the Fe lines, which are available for all SNe of the sample and are considered sufficiently good tracers of the photospheric velocity.\par  


For three SNe of the sample (namely, SNe 1987A, 2009E and OGLE073), the values of $E$, $M_{ej}$, and $R$ have been already estimated independently through the hydrodynamical modelling of their main observables (see, respectively, \citealt{orlando15}, \citealt{pasto12}, and \citealt{terreran16}), using the same code adopted to calculate the grid of models described in Section \ref{models}. Thus we choose these objects as reference SNe in this paper. Moreover they can be used to retrieve information on the accuracy of the relations of sets (\ref{setArnett}), (\ref{setPopov}) and (\ref{setPopov_phot}), by comparing the values derived through the hydrodynamical modelling with those estimated by means of the scaling relations. In Table \ref{tab_parameters_modelling}, we report the values of $E$, $M_{ej}$, and $R$ estimated through procedures of hydrodynamical modelling for the above mentioned three SNe, including also objects modelled with different hydrodynamical codes or semi-analytical approaches.\par

\begin{table*}
   \begin{center}   
   \caption{Same as Table~\ref{tab_observations}, but for the $^{56}$Ni mass inferred from the observed late-time light curve, and the values of $E$, $M_{ej}$, and $R$ estimated through procedures of hydrodynamical modelling from literature (see text for details). Reference to the paper where the modelling was presented is reported in the last column. Masses are in solar units, energy in foe, and progenitor radius in $10^{12}cm$. Estimated uncertainties, when available, are in brackets. $N.A.$ stands for ``not available'' and the symbol $\leq$ indicates upper limit for the $^{56}$Ni mass.}
   \begin{tabular}{llllll}
   \hline\hline
   SN & \multicolumn{1}{c}{$^{56}$Ni mass} & \multicolumn{1}{c}{$E$} & \multicolumn{1}{c}{$M_{ej}$} & \multicolumn{1}{c}{$R$} & \multicolumn{1}{c}{Ref.}\\
      & \multicolumn{1}{c}{[\msun]} & \multicolumn{1}{c}{[foe]} & \multicolumn{1}{c}{[\msun]} & \multicolumn{1}{c}{[$10^{12}cm$]} &\\    
   \hline                                        
   OGLE073$^a$     & $0.47\,(\pm 0.02)$           & $12.4\,(^{+13.0}_{-5.9})$ & $60\,(^{+42}_{-16})$ & $38.0\,(^{+8.0}_{-10.0})$& \citet[][]{terreran16} \\
   2004ek         & $0.217\,(\pm 0.022)^b$       & $N.A.$                    & $N.A.$               & $\simeq 160$             & \citet[][]{taddia16}   \\
   PTF12kso       & $0.230\,(\pm 0.023)^b$       & $N.A.$                    & $N.A.$               & $N.A.$                   &                        \\
   PTF12gcx       & $\leq 0.181\,(\pm 0.018)^b$  & $N.A.$                    & $N.A.$               & $\simeq 24$              & \citet[][]{taddia16}   \\
   2004em         & $0.102\,(\pm 0.010)^b$       & $N.A.$                    & $N.A.$               & $\simeq 22$              & \citet[][]{taddia16}   \\
   2006V          & $0.127\,(\pm 0.013)^b$       & $ 2.4$                    & $17$                 & $5.2$                    & \citet[][]{taddia12}   \\
   2006au         & $\leq 0.073\,(\pm 0.007)^b$  & $ 3.2$                    & $19.3$               & $6.3$                    & \citet[][]{taddia12}   \\
   1998A          & $0.11\,(\pm 0.01)^b$         & $ 5.6$                    & $22$                 & $\lesssim 6$             & \citet[][]{pasto05}    \\
   1987A          & $0.075\,(\pm 0.005)$         & $ 1.3\,(\pm 0.1)$         & $16\,(\pm 1)$        & $3\,(\pm 0.9)$           & \citet[][]{orlando15}  \\
                  &                              & $ 1.1\,(\pm 0.3)$         & $14$                 & $3.4\,(\pm 0.6)$         & \citet[][]{blinnikov00}\\
                  &                              & $ 1.5\,(\pm 0.4)$         & $15\,(\pm 2.2)$      & $6.2\,(\pm 0.9)$         & \citet[][]{zampieri07} \\
   2000cb         & $0.083\,(\pm 0.039)$         & $ 4.4\,(\pm 0.3)$         & $22.3\,(\pm 1.0)$    & $2.4\,(\pm 0.96)$        & \citet[][]{uc11}       \\
                  &                              & $ 2$                      & $17.5$               & $3$                      & \citet[][]{kleiser11}  \\                 
   2005ci         & $0.065\,(\pm 0.006)^b$       & $ 1$                      & $20$                 & $\simeq 2.3$             & \citet[][]{taddia16}   \\
   2009mw         & $0.062\,(\pm 0.006)^b$       & $ 1.0$                    & $17.5$               & $\simeq 2.1$             & \citet[][]{takats16}   \\
   2009E          & $0.040\,(^{+0.015}_{-0.011})$& $ 0.6\,(\pm 0.2)$         & $19\,(\pm 2.8)$      & $7\,(\pm 1.0)$           & \citet[][]{pasto12}    \\
   DES16C3cje$^c$ & $0.068\,(\pm 0.007)^b$       & $ 0.11$                   & $15$                 & $\simeq 57.7$            & \citet[][]{gutierrez20}\\
                  &                              & $ 1$                      & $40$                 & $\simeq 7.0$             & \citet[][]{gutierrez20}\\
  \hline                                   
 \end{tabular}
 \label{tab_parameters_modelling}
\end{center} 
Note that \citet[][]{terreran16}, \citet[][]{orlando15} and \citet[][]{pasto12} used the radiation-hydrodynamics code presented in \citet[][]{pumo10} and \citet[][]{PZ11} (see text for details); \citet[][]{taddia16} applied a relation estimated from a series of hydrodynamical models calculated with the SuperNova Explosion Code \citep[SNEC;][]{morozowa15} coupled with the Modules for Experiments in Stellar Astrophysics \citep[MESA;][]{paxton11}; \citet[][]{taddia12} used the semi-analytic model of \citet[][]{IP92}; \citet[][]{pasto05} and \citet[][]{zampieri07} used different versions of the semi-analytic model presented in \citet[][]{zampieri03}; \citet[][]{blinnikov00} used the hydrodynamics code STELLA \citep[][]{BB93,blinnikov98}; \citet[][]{kleiser11} used the radiation-hydrodynamics code presented in \citet[][]{young04}; \citet[][]{uc11} used their own hydrodynamical model; \citet[][]{takats16} and \citet[][]{gutierrez20} used the radiation-hydrodynamics code presented in \citet*[][]{bertsen11}.\\
$^a$ The $^{56}$Ni mass inferred from the observations and the calculated values of $E$, $M_{ej}$, and $R$ were estimated considering that the explosion of OGLE073's progenitor occurred only one day before discovery. Assuming that the explosion occurred $\sim 90$ days before the discovery, the $^{56}$Ni mass could be as high as $\sim 1.1$ \msun, and the values of $E$ and $M_{ej}$ should further increase \citep[see][for further details]{terreran16}.\\  
$^b$ Value estimated considering that the uncertainties in distance and explosion epoch lead typically to an error in the $^{56}$Ni mass of the order of 10\% \citep[see][]{taddia16}.\\
$^c$ \citet[][]{gutierrez20} are not able to disentangle between two alternative scenarios to explain DES16C3cje. The reported values of $E$, $M_{ej}$, and $R$ refer to such scenarios. In both cases an additional energy input (compared to what expected from standard powering by radioactive decay of $^{56}$Ni) is necessary to explain the late-time light curve of DES16C3cje, and the $^{56}$Ni mass considered in the hydrodynamical models (0.075\Msun and 0.08\Msun for the model having $E$ of 0.11 and 1 foe, respectively) is slightly higher than the value inferred from the late-time observations by comparing the bolometric light curve of DES16C3cje to that of SN 1987A.\\
\end{table*}                                                                                                                                                                                                                                                                                

\section{Results and discussion}
\label{results}

\subsection{Scaling relations}
\label{scaling_rel}

Tables \ref{tab_Arnett}, \ref{tab_Popov}, and \ref{tab_Popov_phot} show the values of $E$, $M_{ej}$, and $R$ derived from relations of sets (\ref{setArnett}), (\ref{setPopov}), and (\ref{setPopov_phot}), respectively. For the values derived from relations of set (\ref{setArnett}), we consider three different reference SNe (namely, SNe 1987A, 2009E and OGLE073). For the values derived from relations of sets (\ref{setPopov}) and (\ref{setPopov_phot}), we consider a reduced sample of SN 1987A-like objects and only one reference SN (namely, SN 1987A) because the determinations of $L_m$ is not always possible (cf.~Table~\ref{tab_observations}).\par 

\begin{figure*}
 \includegraphics[angle=-90,width=160mm]{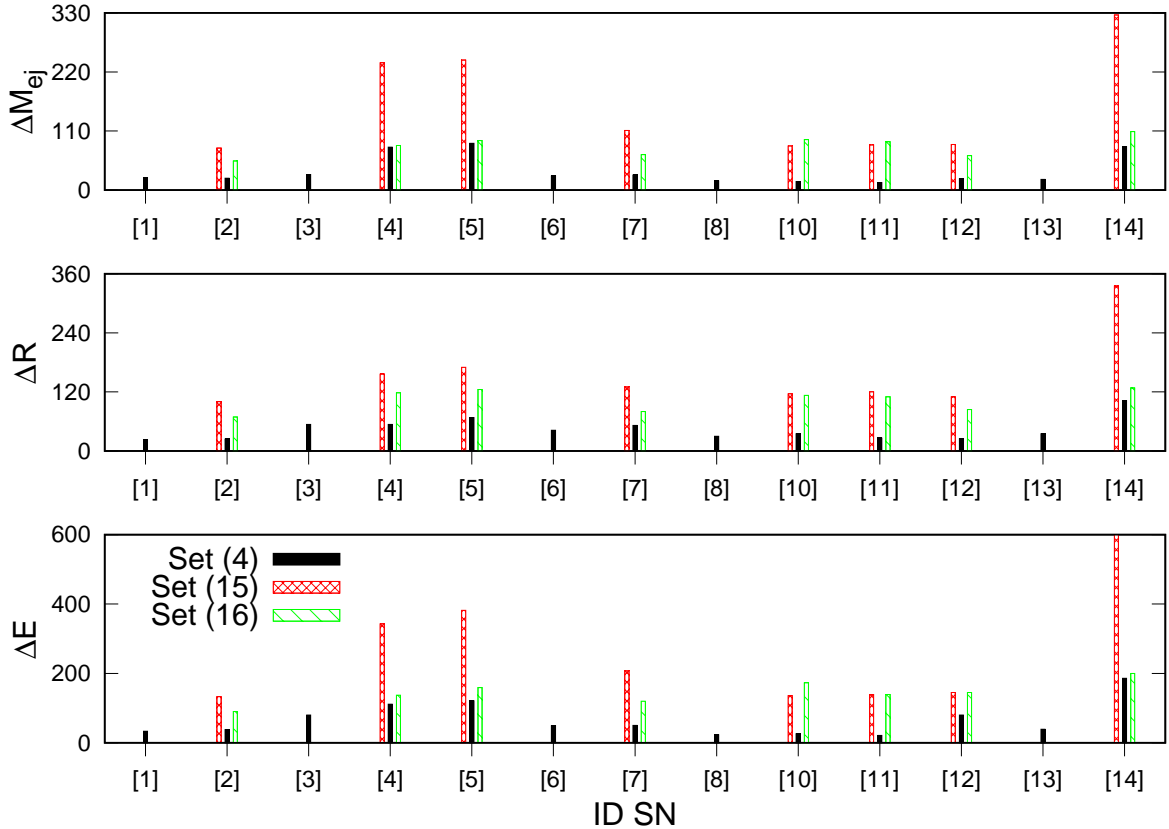} 
 \caption{ {Relative percentage errors on the values of $M_{ej}$ (top panel), $R$ (middle panel), and $E$ (bottom panel) derived from the scaling relations of sets (\ref{setArnett}) (filled black boxes), (\ref{setPopov}) (double-dashed red boxes), and (\ref{setPopov_phot}) (dashed green boxes) for the sample of SN 1987A-like objects, using SN 1987A as reference (cf.~Tables \ref{tab_Arnett} to \ref{tab_Popov_phot}).``ID SN'' identifies an individual SN 1987A-like object in the sample (namely, [1] $\rightarrow$ OGLE073, [2] $\rightarrow$ 2004ek, [3] $\rightarrow$ PTF12kso, [4] $\rightarrow$ PTF12gcx, [5] $\rightarrow$ 2004em, [6] $\rightarrow$ 2006V, [7] $\rightarrow$ 2006au, [8] $\rightarrow$ 1998A, [10] $\rightarrow$ 2000cb, [11] $\rightarrow$ 2005ci, [12] $\rightarrow$ 2009mw, [13] $\rightarrow$ 2009E, [14] $\rightarrow$ DES16C3cje).}
 \label{figErrRel}}
\end{figure*}

The data reported in Tables \ref{tab_Arnett} to \ref{tab_Popov_phot} (and also in Appendix \ref{app_scaling_rel}) indicate that the most precise relations (i.e.~characterised by lower relative errors) are those of set (\ref{setArnett}). Indeed, as clearly highlighted in Figure \ref{figErrRel}, relative errors on the values of $E$, $M_{ej}$, and $R$  derived from this set of scaling relations are lower compared to those obtained when using the relations of sets (\ref{setPopov}) and (\ref{setPopov_phot}). This is expected given that the relations of set (\ref{setArnett}) are power laws depending on less parameters and with smaller exponents with respect to relations of sets (\ref{setPopov}) and (\ref{setPopov_phot}). Consequently, errors on the parameters $L_m$, $L_M$, $v_M$, and $t_M$ propagate to a less extent.\par

\begin{table*}
   \centering
   \caption{Values of $E$, $M_{ej}$, and $R$ derived from relations of set (\ref{setArnett}), considering three different reference SNe (namely, SNe 1987A, 2009E and OGLE073). Masses are in solar units, progenitor radius in $10^{12}cm$, and energy in foe. Estimated uncertainties on the inferred parameters are in round brackets. The adopted values of $E$, $M_{ej}$, and $R$ for the reference SNe are put between square brackets (see the text for details and cf.~Table~\ref{tab_parameters_modelling}).}
   \addtolength{\tabcolsep}{-4.0pt}     
   \begin{tabular}{llllllllll}
   \hline\hline
   SN & \multicolumn{1}{c}{$E$} & \multicolumn{1}{c}{$M_{ej}$} & \multicolumn{1}{c}{$R$} & \multicolumn{1}{c}{$E$} & \multicolumn{1}{c}{$M_{ej}$} & \multicolumn{1}{c}{$R$} & \multicolumn{1}{c}{$E$} & \multicolumn{1}{c}{$M_{ej}$} & \multicolumn{1}{c}{$R$} \\
      & \multicolumn{1}{c}{[foe]} & \multicolumn{1}{c}{[\msun]} & \multicolumn{1}{c}{[$10^{12}cm$]} & \multicolumn{1}{c}{[foe]} & \multicolumn{1}{c}{[\msun]} & \multicolumn{1}{c}{[$10^{12}cm$]} & \multicolumn{1}{c}{[foe]} & \multicolumn{1}{c}{[\msun]} & \multicolumn{1}{c}{[$10^{12}cm$]} \\
   \hline   
      & \multicolumn{3}{c}{ref.~SN: 1987A} & \multicolumn{3}{c}{ref.~SN: 2009E} & \multicolumn{3}{c}{ref.~SN: OGLE073} \\    
   \hline
   OGLE073    & $26.6\,(\pm33\%)$  & $52.8\,(\pm23\%)$ & $5.1\,(\pm23\%)$  & $23.9\,(\pm45\%)$  & $71.0\,(\pm24\%)$ & $10.0\,(\pm44\%)$  &[$12.4\,(^{+13.0}_{-5.9})$ & $60\,(^{+42}_{-16})$ & $38.0\,(^{+8.0}_{-10.0})$]\\
   2004ek     & $13.4\,(\pm38\%)$  & $30.3\,(\pm22\%)$ & $3.1\,(\pm24\%)$  & $12.0\,(\pm48\%)$  & $40.7\,(\pm23\%)$ & $ 6.3\,(\pm43\%)$  & $ 6.2\,(\pm64\%)$         & $34.4\,(\pm42\%)$    & $22.9\,(\pm26\%)$         \\
   PTF12kso   & $ 5.9\,(\pm80\%)$  & $17.6\,(\pm29\%)$ & $2.4\,(\pm54\%)$  & $ 5.3\,(\pm85\%)$  & $23.6\,(\pm30\%)$ & $ 5.0\,(\pm64\%)$  & $ 2.7\,(\pm96\%)$         & $19.9\,(\pm46\%)$    & $18.0\,(\pm55\%)$         \\
   PTF12gcx   & $ 5.2\,(\pm111\%)$ & $15.1\,(\pm80\%)$ & $1.7\,(\pm54\%)$  & $ 4.6\,(\pm115\%)$ & $20.3\,(\pm88\%)$ & $ 3.4\,(\pm64\%)$  & $ 2.4\,(\pm122\%)$        & $17,1\,(\pm95\%)$    & $12.3\,(\pm55\%)$         \\
   2004em     & $10.4\,(\pm121\%)$ & $32.2\,(\pm87\%)$ & $1.6\,(\pm68\%)$  & $ 9.3\,(\pm125\%)$ & $43.2\,(\pm87\%)$ & $ 3.2\,(\pm77\%)$  & $ 4.8\,(\pm132\%)$        & $36.5\,(\pm94\%)$    & $11.5\,(\pm69\%)$         \\
   2006V      & $ 2.7\,(\pm49\%)$  & $17.4\,(\pm26\%)$ & $2.9\,(\pm42\%)$  & $ 2.4\,(\pm58\%)$  & $23.3\,(\pm27\%)$ & $ 6.0\,(\pm55\%)$  & $ 1.2\,(\pm72\%)$         & $19.7\,(\pm44\%)$    & $21.7\,(\pm43\%)$         \\
   2006au     & $ 8.5\,(\pm50\%)$  & $23.6\,(\pm28\%)$ & $1.1\,(\pm52\%)$  & $ 7.6\,(\pm77\%)$  & $31.6\,(\pm30\%)$ & $ 2.3\,(\pm62\%)$  & $ 4.0\,(\pm89\%)$         & $26.7\,(\pm45\%)$    & $ 8.2\,(\pm52\%)$         \\
   1998A      & $ 3.9\,(\pm23\%)$  & $25.4\,(\pm17\%)$ & $1.8\,(\pm30\%)$  & $ 3.5\,(\pm37\%)$  & $34.1\,(\pm18\%)$ & $ 3.8\,(\pm47\%)$  & $ 1.8\,(\pm57\%)$         & $28.8\,(\pm40\%)$    & $13.7\,(\pm32\%)$         \\
   1987A      &[$ 1.3\,(\pm 0.1)$  & $16\,(\pm 1)$     & $3\,(\pm 0.9)$]   & $ 1.2\,(\pm40\%)$  & $21.5\,(\pm19\%)$ & $ 6.1\,(\pm41\%)$  & $ 0.6\,(\pm58\%)$         & $18.2\,(\pm40\%)$    & $22.1\,(\pm22\%)$         \\
   2000cb     & $ 6.6\,(\pm26\%)$  & $19.3\,(\pm16\%)$ & $0.6\,(\pm35\%)$  & $ 5.9\,(\pm40\%)$  & $26.0\,(\pm18\%)$ & $ 1.3\,(\pm50\%)$  & $ 3.0\,(\pm59\%)$         & $22.0\,(\pm39\%)$    & $ 4.7\,(\pm37\%)$         \\
   2005ci     & $ 3.3\,(\pm21\%)$  & $23.1\,(\pm14\%)$ & $1.2\,(\pm26\%)$  & $ 2.9\,(\pm58\%)$  & $31.1\,(\pm16\%)$ & $ 2.5\,(\pm44\%)$  & $ 1.5\,(\pm28\%)$         & $26.2\,(\pm39\%)$    & $ 9.1\,(\pm28\%)$         \\
   2009mw     & $ 4.6\,(\pm80\%)$  & $21.7\,(\pm21\%)$ & $1.0\,(\pm25\%)$  & $ 3.3\,(\pm48\%)$  & $29.2\,(\pm22\%)$ & $ 2.1\,(\pm43\%)$  & $ 1.7\,(\pm64\%)$         & $24.7\,(\pm42\%)$    & $ 7.5\,(\pm27\%)$         \\
   2009E      & $ 0.7\,(\pm39\%)$  & $14.1\,(\pm20\%)$ & $3.4\,(\pm35\%)$  &[$ 0.6\,(\pm 0.2)$  & $19\,(\pm 2.8)$   & $7\,(\pm 1.0)$]    & $ 0.3\,(\pm67\%)$         & $16.1\,(\pm41\%)$    & $25.3\,(\pm37\%)$         \\
   DES16C3cje & $ 0.6\,(\pm186\%)$ & $23.2\,(\pm81\%)$ & $4.7\,(\pm102\%)$ & $ 0.5\,(\pm238\%)$ & $31.1\,(\pm81\%)$ & $ 9.6\,(\pm165\%)$ & $ 0.3\,(\pm240\%)$        & $26.3\,(\pm89\%)$    & $34.7\,(\pm37\%)$         \\
   \hline                                   
 \end{tabular}
 \label{tab_Arnett}
\end{table*}

\begin{table}
   \centering
   \caption{Same as Table~\ref{tab_Arnett}, but for values of $E$, $M_{ej}$, and $R$ derived from relations of set (\ref{setPopov}). The symbol $\leq$ ($\geq$) indicates that it is possible to estimate just an upper (lower) limit for the considered physical quantity (see the text for details and cf.~Table~\ref{tab_observations}).}
   \addtolength{\tabcolsep}{-3.2pt} 
   \begin{tabular}{llllllllll}
   \hline\hline
   SN & \multicolumn{1}{c}{$E$}   & \multicolumn{1}{c}{$M_{ej}$} & \multicolumn{1}{c}{$R$} \\
      & \multicolumn{1}{c}{[foe]} & \multicolumn{1}{c}{[\msun]}  & \multicolumn{1}{c}{[$10^{12}$\,cm]}\\
   \hline   
      & \multicolumn{3}{c}{ref.~SN: 1987A}\\    
   \hline
   2004ek     & $ 0.2\,(\pm133\%)$      & $ 2.1\,(\pm78\%)$      & $87.6\,(\pm100\%)   $  \\
   PTF12gcx   & $ 0.7\,(\pm343\%)$      & $ 4.5\,(\pm237\%)$     & $ 7.0\,(\pm156\%)   $  \\
   2004em     & $ 2.7\,(\pm382\%)$      & $14.3\,(\pm242\%)$     & $ 8.9\,(\pm170\%)   $  \\
   2006au     & $ 1.9\,(\pm207\%)$      & $ 9.5\,(\pm111\%)$     & $ 7.0\,(\pm130\%)   $  \\
   1987A      &[$ 1.3\,(\pm 0.1)$       & $16\,(\pm 1)$          & $3\,(\pm 0.9)       $] \\
   2000cb     & $\geq 74.4\,(\pm136\%)$ & $\geq 83.0\,(\pm82\%)$ & $\leq 0.2\,(\pm116\%)$ \\
   2005ci     & $\geq 29.3\,(\pm139\%)$ & $\geq 86.1\,(\pm84\%)$ & $\leq 0.6\,(\pm120\%)$ \\
   2009mw     & $\geq 10.6\,(\pm145\%)$ & $\geq 40.8\,(\pm85\%)$ & $\leq 1.2\,(\pm110\%)$ \\
   DES16C3cje & $ 0.4\,(\pm600\%)$      & $18.4\,(\pm326\%)$     & $10.6\,(\pm335\%)$     \\
   \hline                                   
 \end{tabular}
 \label{tab_Popov}
\end{table}

\begin{table}
   \centering
   \caption{Same as Table~\ref{tab_Popov}, but for values of $E$, $M_{ej}$, and $R$ derived from relations of set (\ref{setPopov_phot}). }
   \addtolength{\tabcolsep}{-3.2pt}  
   \begin{tabular}{llllllllll}
   \hline\hline
   SN & \multicolumn{1}{c}{$E$}   & \multicolumn{1}{c}{$M_{ej}$} & \multicolumn{1}{c}{$R$} \\
      & \multicolumn{1}{c}{[foe]} & \multicolumn{1}{c}{[\msun]}  & \multicolumn{1}{c}{[$10^{12}cm$]}\\
   \hline   
      & \multicolumn{3}{c}{ref.~SN: 1987A}\\    
   \hline
   2004ek     & $ 0.4\,(\pm 90\%)$      & $ 3.3\,(\pm54\%)$      & $54.0\,(\pm69\%)    $  \\
   PTF12gcx   & $ 1.5\,(\pm137\%)$      & $ 6.6\,(\pm83\%)$      & $ 4.8\,(\pm118\%)   $  \\
   2004em     & $ 0.2\,(\pm159\%)$      & $ 3.8\,(\pm92\%)$      & $33.5\,(\pm125\%)   $  \\
   2006au     & $ 0.1\,(\pm120\%)$      & $ 2.6\,(\pm66\%)$      & $25.1\,(\pm80\%)    $  \\
   1987A      &[$ 1.3\,(\pm 0.1)$       & $16\,(\pm 1)$          & $3\,(\pm 0.9)       $] \\
   2000cb     & $\geq 1.2\,(\pm173\%)$  & $\geq 10.5\,(\pm94\%)$ & $\leq 2.0\,(\pm113\%)$ \\
   2005ci     & $\geq 0.6\,(\pm139\%)$  & $\geq 12.0\,(\pm90\%)$ & $\leq 4.1\,(\pm110\%)$ \\
   2009mw     & $\geq 0.2\,(\pm145\%)$  & $\geq 5.3\,(\pm64\%)$  & $\leq 9.3\,(\pm 84\%)$ \\
   DES16C3cje & $ 0.1\,(\pm200\%)$      & $ 6.6\,(\pm109\%)$     & $29.4\,(\pm128\%)$     \\
   \hline                                   
 \end{tabular}
 \label{tab_Popov_phot}
\end{table}


\begin{figure*}
 \includegraphics[angle=-90,width=160mm]{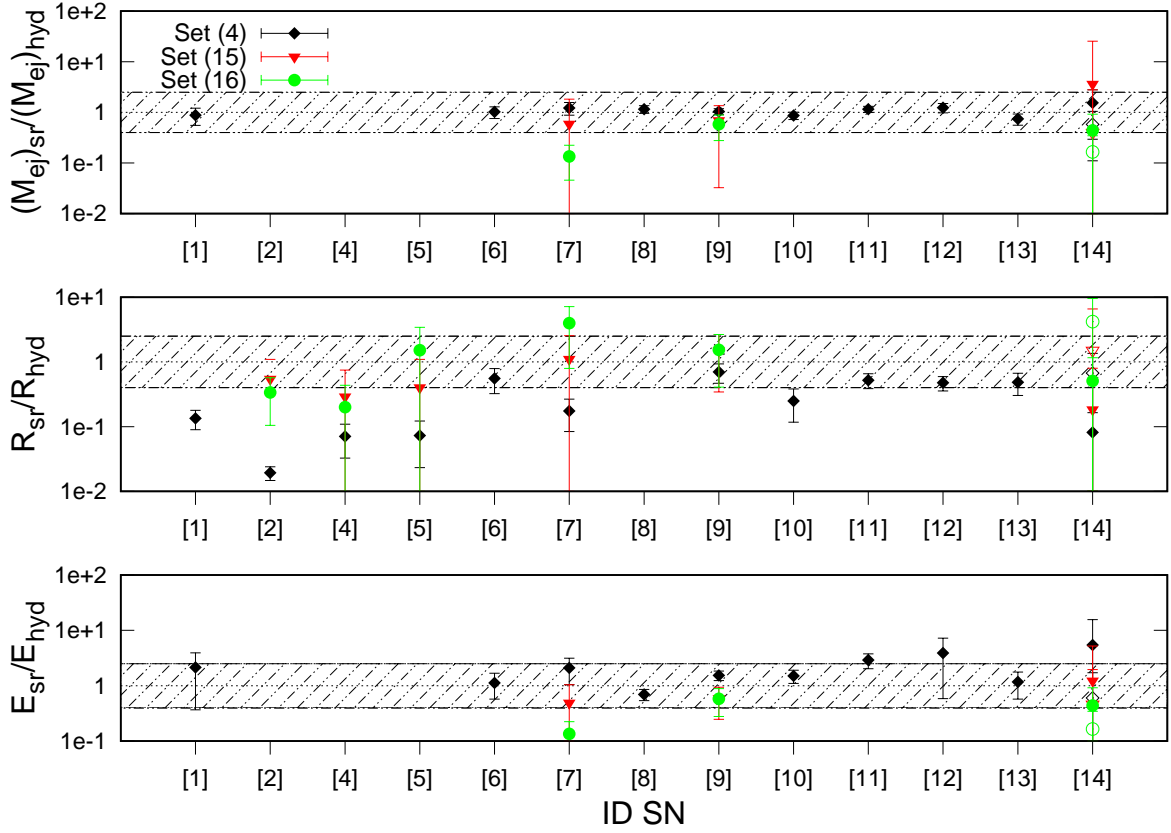} 
 \caption{ {Top panel: ratio of the value of $M_{ej}$ derived from the scaling relations to that estimated through hydrodynamical modelling, $(M_{ej})_{sr}$/$(M_{ej})_{hyd}$ for the sample of SN 1987A-like objects, using SN 1987A as reference (except for SN 1987A, for which we use model 5 -- cf.~Section \ref{models} --- as reference). As in Figure \ref{figErrRel}, ``ID SN'' identifies the SN 1987A-like objects (namely, [1] $\rightarrow$ OGLE073, [2] $\rightarrow$ 2004ek, [4] $\rightarrow$ PTF12gcx, [5] $\rightarrow$ 2004em, [6] $\rightarrow$ 2006V, [7] $\rightarrow$ 2006au, [8] $\rightarrow$ 1998A, [9] $\rightarrow$ 1987A, [10] $\rightarrow$ 2000cb, [11] $\rightarrow$ 2005ci, [12] $\rightarrow$ 2009mw, [13] $\rightarrow$ 2009E, [14] $\rightarrow$ DES16C3cje). The adopted values of $(M_{ej})_{hyd}$ are reported in Table \ref{tab_parameters_modelling}. In particular, for SNe 1987A and 2000cb we adopt the values reported in \citet{orlando15} and \citet{uc11}, respectively. For SN DES16C3cje, we consider both the explosive scenarios presented in \citet{gutierrez20} [open symbols refer to the second scenario characterized by an higher values of $(M_{ej})_{hyd}$, cf.~last line in Table \ref{tab_parameters_modelling}]. The value of $(M_{ej})_{hyd}$ and, consequently, the ratio $(M_{ej})_{sr}$/$(M_{ej})_{hyd}$, are evaluated using the first relation of the sets (\ref{setArnett}) (black diamonds), (\ref{setPopov}) (red triangles), and (\ref{setPopov_phot}) (green circles). The dashed area constrains the $0.4 \leq (M_{ej})_{sr}/(M_{ej})_{hyd} \leq 2.5$ range. SNe whose values of $(M_{ej})_{sr}$ and/or $(M_{ej})_{hyd}$ are upper limits, lower limits or not available (cf.~Tables \ref{tab_parameters_modelling} to \ref{tab_Popov_phot}), are not reported. Middle panel: as top panel, but for the $R_{sr}$/$R_{hyd}$ ratio between the value of $R$ derived from the scaling relations and that estimated with hydrodynamical modelling. Bottom panel: as top panel, but for the $E_{sr}$/$E_{hyd}$ ratio between the value of $E$ derived from the scaling relations and that estimated with hydrodynamical modelling.}
 \label{figAccuracy}}
\end{figure*}

\begin{table}
   \centering
   \caption{Selected parameters (see text for details) for the grid of radiation-hydrodynamical models considered in this paper (cf.~Section \ref{models}). Luminosities are in $10^{40}$~erg~s$^{-1}$, velocity in km~s$^{-1}$, and time in days.}
   \addtolength{\tabcolsep}{-1.3pt}  
   \begin{tabular}{lllll}
   \hline\hline
   Model & \multicolumn{1}{c}{$L_m$} & \multicolumn{1}{c}{$L_M$} & \multicolumn{1}{c}{$v_M$} & \multicolumn{1}{c}{$t_M$}\\
         & \multicolumn{1}{c}{[$10^{40}$~erg s$^{-1}$]} & \multicolumn{1}{c}{[$10^{40}$~erg s$^{-1}$]} & \multicolumn{1}{c}{[km s$^{-1}$]} & \multicolumn{1}{c}{[d]}\\    
   \hline                                        
   $1$   & $7.8$  & $20.0$  & $2537$  & $41.7$ \\
   $2$   & $7.8$  & $20.2$  & $2380$  & $44.2$ \\
   $3$   & $7.8$  & $23.7$  & $2472$  & $44.9$ \\
   $4$   & $7.9$  & $43.8$  & $1368$  & $91.7$ \\
   $5$   & $7.3$  & $63.8$  & $1418$  & $105.1$\\ 
   $6$   & $8.0$  & $88.0$  & $1698$  & $105.3$\\
   $7$   & $8.3$  & $213.4$ & $1403$  & $137.8$\\
   $8$   & $8.4$  & $417.0$ & $1584$  & $154.2$\\
  \hline                                   
 \end{tabular}
 \label{tab_parameters_grid}
\end{table}

\begin{figure}
 \includegraphics[angle=-90,width=85mm]{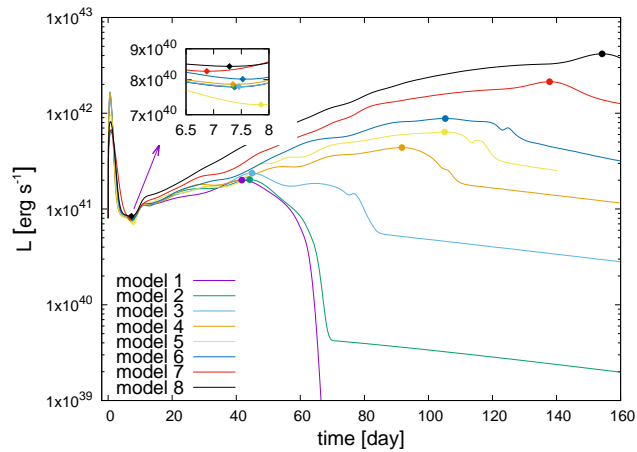} 
 \caption{Bolometric luminosities during the first 160 days after the explosion for the grid of hydrodynamical computations considered in this work (cf.~Section \ref{models}). The solid diamond and solid circle symbols mark the location of the point of ($L_m$, $t_m$) and ($L_M$, $t_M$) coordinates, respectively. The inset box is a zoom of the bolometric luminosities in the range 6.5-8 days after the explosion.
 \label{figBolMod}}
\end{figure}
\begin{figure}
 \includegraphics[angle=-90,width=85mm]{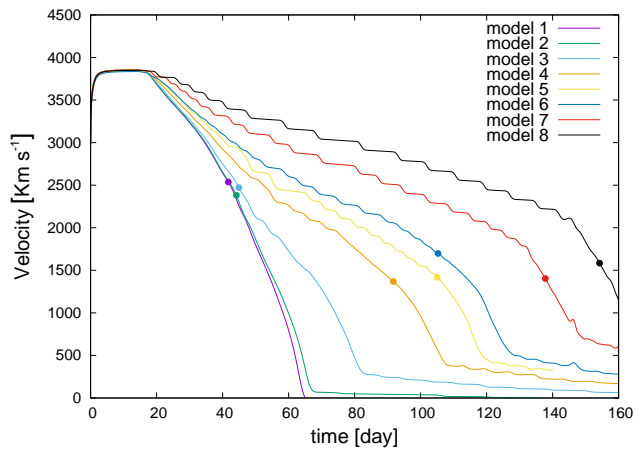} 
 \caption{Same as Figure \ref{figBolMod}, but for the behavior of the photospheric velocity. The solid circle symbols mark the location of the point of ($v_M$, $t_M$) coordinates.
 \label{figVelMod}}
\end{figure}
\begin{figure*}
 \includegraphics[angle=-90,width=160mm]{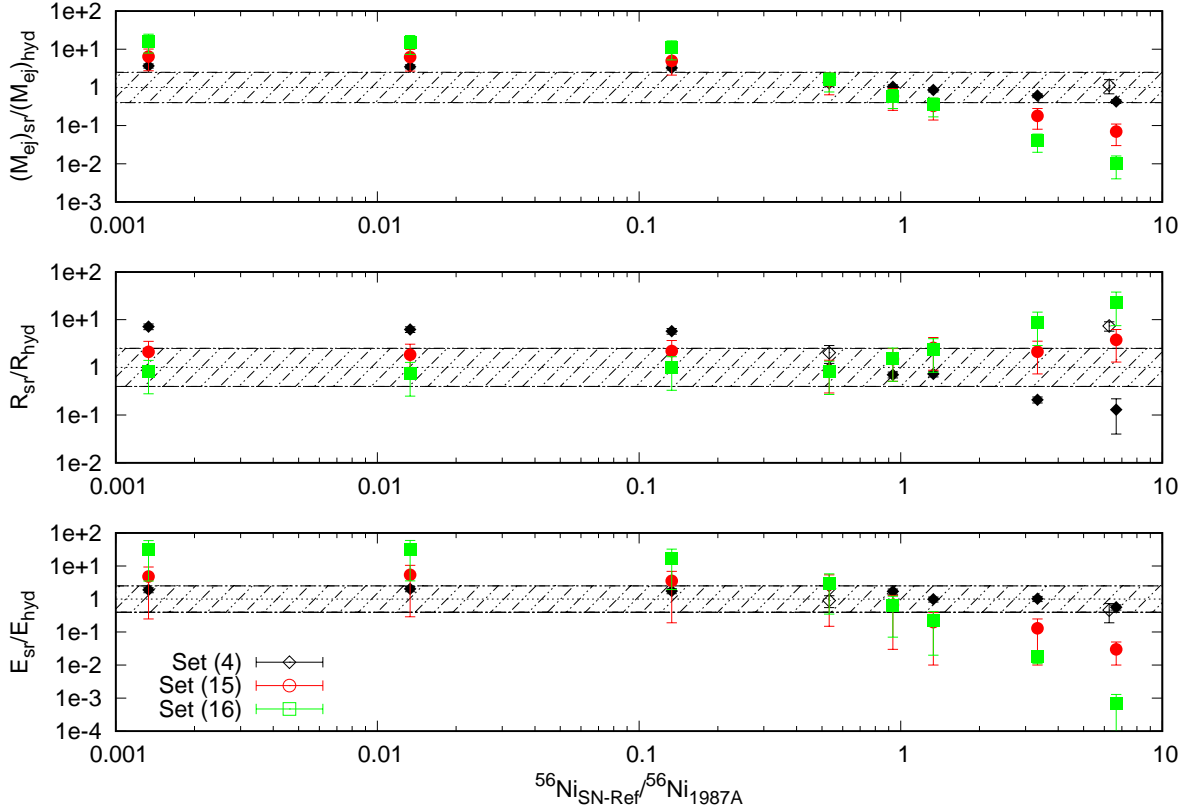}
\caption{Top panel: $(M_{ej})_{sr}$/$(M_{ej})_{hyd}$ ratio as a function of the amount of $^{56}$Ni of the reference SN, $^{56}$Ni$_{SN-Ref}$, normalized to the amount of the $^{56}$Ni of the SN 1987A at which the scaling relations are applied. The value of $E_{hyd}$ is taken from \citet{orlando15} (cf.~Section \ref{observations} and Table \ref{tab_parameters_modelling}). The value of $(M_{ej})_{sr}$ and, consequently, the ratio $(M_{ej})_{sr}$/$(M_{ej})_{hyd}$ are evaluated using the first relation of the sets (\ref{setArnett}) (black diamonds), (\ref{setPopov}) (red cirles), and (\ref{setPopov_phot}) (green squares). Open and filled symbols, respectively, refer to the usage of real (cf.~Section \ref{observations}) or simulated (cf.~Section \ref{models}) reference SNe. The dashed area constrains the $0.4 \leq (M_{ej})_{sr}$/$(M_{ej})_{hyd} \leq 2.5$ range. Middel panel: as top panel, but for the $R_{sr}$/$R_{hyd}$ ratio. Bottom panel: as top panel, but for the $E_{sr}$/$E_{hyd}$ ratio.
 \label{figratio87A}}
\end{figure*}
\begin{figure}
 \includegraphics[angle=-90,width=85mm]{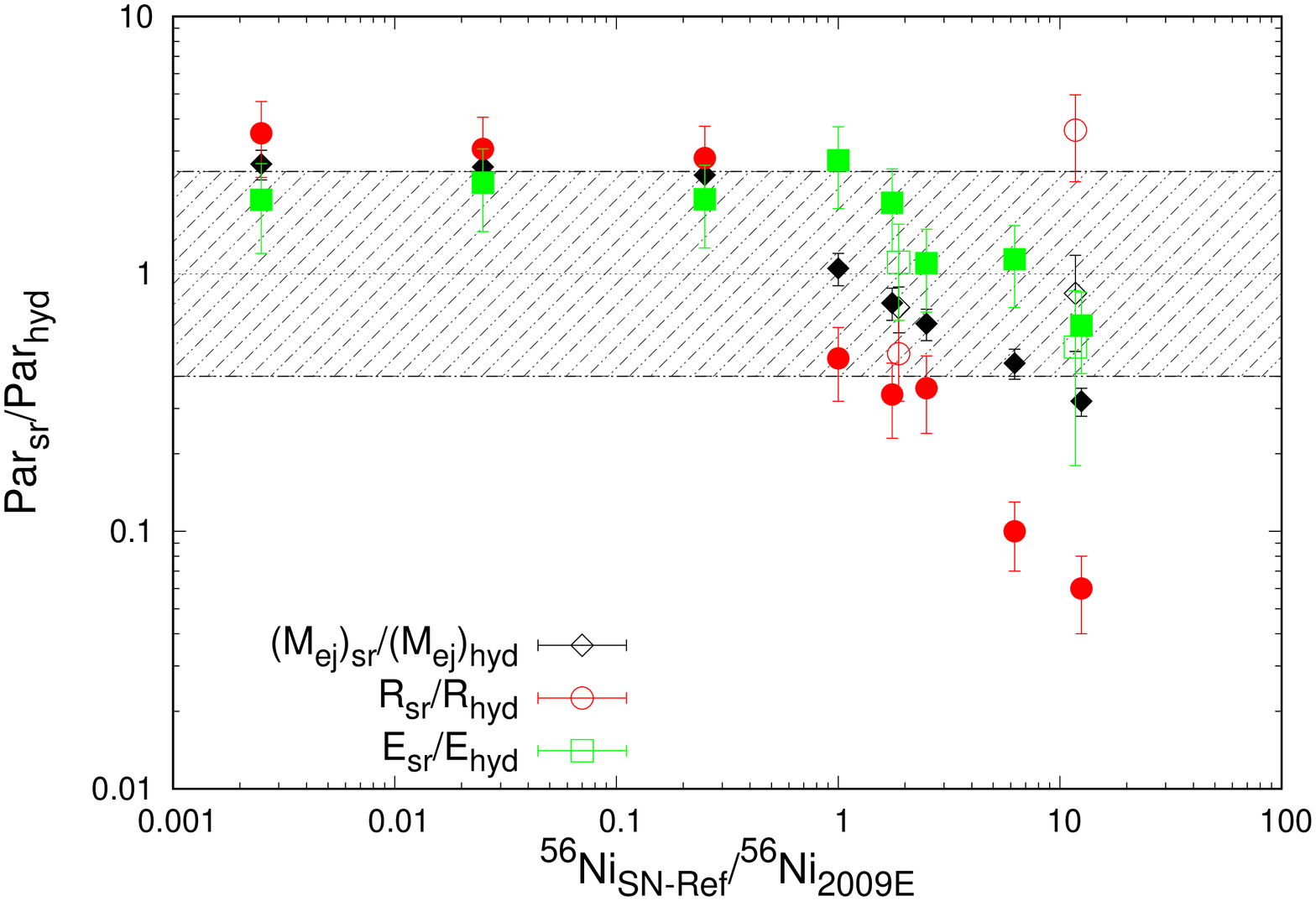} 
 \caption{$(M_{ej})_{sr}$/$(M_{ej})_{hyd}$ (black diamonds), $R_{sr}$/$R_{hyd}$ (red cirles), and $E_{sr}$/$E_{hyd}$ (green squares) ratios, as a function of the amount of $^{56}$Ni of the reference SN, $^{56}$Ni$_{SN-Ref}$, normalized to the amount of the $^{56}$Ni of the SN 2009E at which the scaling relations of set (\ref{setArnett}) are applied. The values of $(M_{ej})_{hyd}$, $R_{hyd}$, and $E_{hyd}$ are taken from \citet{pasto12} (cf.~Section \ref{observations} and Table \ref{tab_parameters_modelling}). Open and filled symbols, respectively, refer to the usage of real (cf.~Section \ref{observations}) and simulated SNe through our hydrodynamical modelling (cf.~Section \ref{models}) as reference SNe. The dashed area constrains the $0.4 \leq Par_{sr}$/$Par_{hyd} \leq 2.5$ range, where $Par$ is equal to one of the three parameters $M_{ej}$, $R$, and $E$.
 \label{figratio09E}}
\end{figure}
\begin{figure}
 \includegraphics[angle=-90,width=85mm]{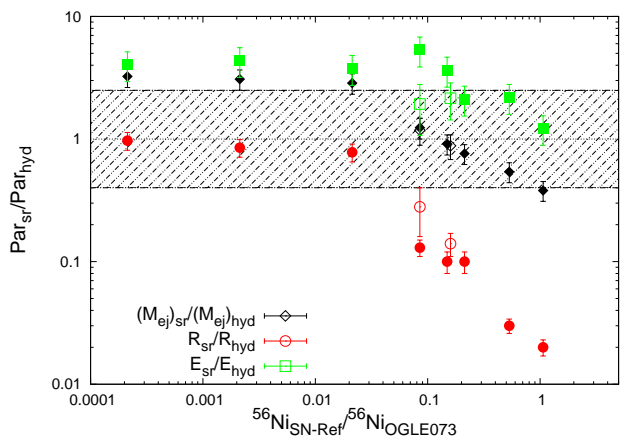} 
 \caption{As Figure \ref{figratio09E}, but for the explosive event OGLE073. The values of $(M_{ej})_{hyd}$, $R_{hyd}$, and $E_{hyd}$ are taken from \citet{terreran16} (cf.~Section \ref{observations} and Table \ref{tab_parameters_modelling}). 
 \label{figratioOGLE073}}
\end{figure}
\begin{figure}
 \includegraphics[angle=-90,width=85mm]{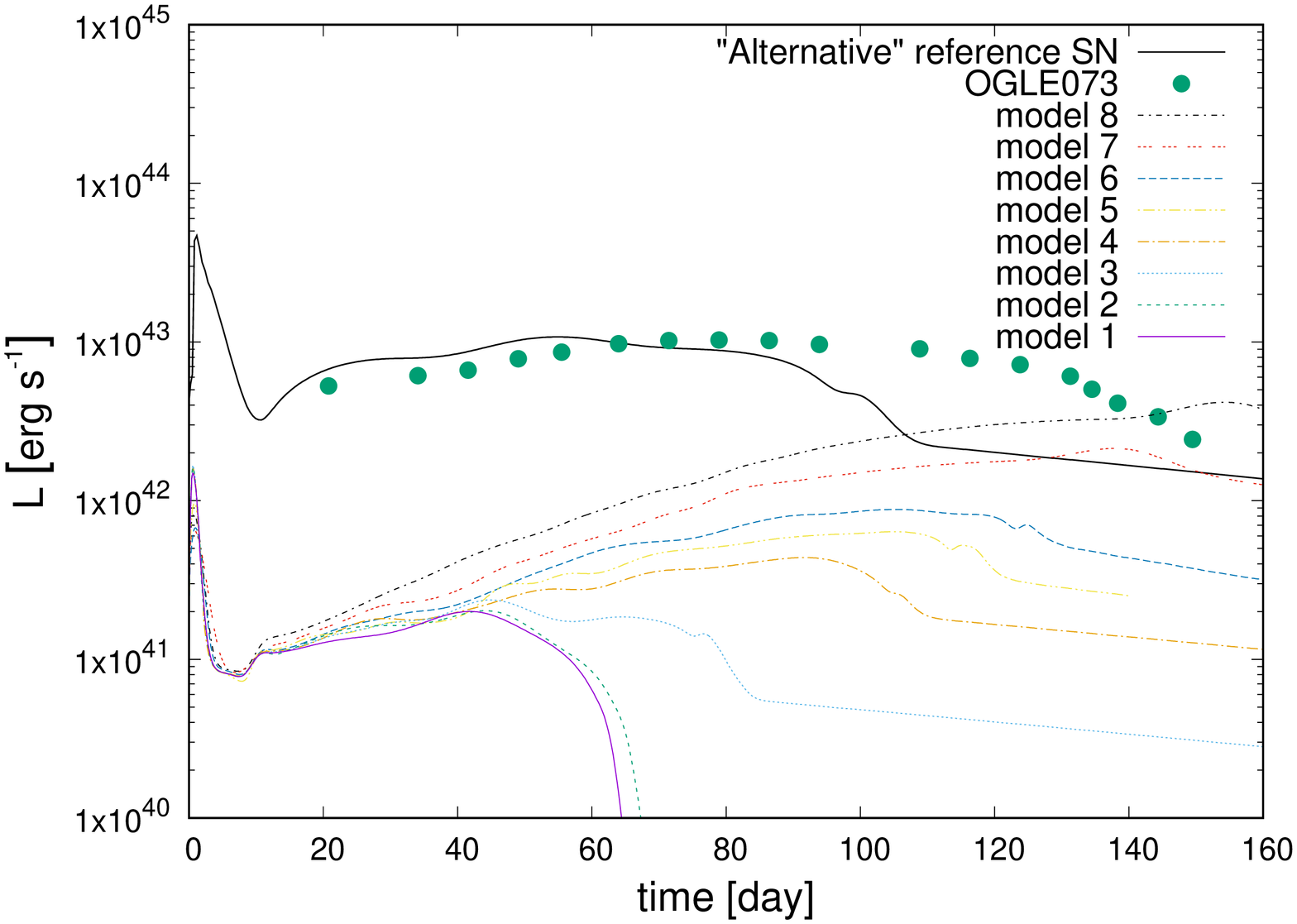} 
 \caption{(Pseudo-)Bolometric luminosity (during the first 160 days after the explosion) of the SN 1987A-like model used as ``alternative'' reference SN for evaluating the $R$ parameter of OGLE073 through the third relation of set (\ref{setArnett}) in a more accurate way (see text for details). It is calculated like the models of Table \ref{tabmodels}, but its parameters are $M_{ej}=45$\msun, $E=11.5$\,foe, $R=40$x$10^{12}$\,cm, and $M_{Ni}=0.52$\,\msun. The bolometric luminosities of OGLE073 (green circles) and for the models of Table \ref{tabmodels} (lines of different styles and pattern) are also reported for sake of comparison.
 \label{figmodelloNew}}
\end{figure}

Furthermore, the data show (see also Figure \ref{figAccuracy}) that the deviation between the values of $E$, $M_{ej}$, and $R$ derived from the scaling relations and those estimated with hydrodynamical or semi-analytical approaches is generally smaller when considering the first two relations of set (\ref{setArnett}) [i.e.~the relations linking $E$ and $M_{ej}$ to $v_M$ and $t_M$] and the third relation of set (\ref{setPopov}) [i.e.~the relation linking $R$ to $L_m$, $t_M$, and $v_M$]. So these three relations appear to be the best ones in terms of accuracy. This is probably due to the different dependence of the values of $L_m$, $L_M$, $v_M$, and $t_M$ from the \chem{56}{Ni} mass in the ejected material. In fact, although all the derived scaling relations [i.e. relations of sets (\ref{setArnett}), (\ref{setPopov}), and (\ref{setPopov_phot})] are based on analytical models that do not consider the heating effects of radioactive isotopes (primarily, the \chem{56}{Ni}) in the ejected material (cf.~Section \ref{modelling}), in real long-rising SNe these effetcs have a non-negligible impact on the bolometric luminosities and photospheric velocity and, consequently, on the above mentioned parameters. However the \chem{56}{Ni} mass affects the various parameters to a different extent \citep[see also e.g.][and references therein]{PZ11,PZ13}. As also confirmed by the behavior of our realistic SN 1987A-like models which are simulated including the heating effects due to the presence of radioactive isotopes (cf.~Section \ref{models}), the parameter which is most affected by the \chem{56}{Ni} mass is $L_M$, while $t_M$ and $v_M$ are less affected by it, and $L_m$ is essentially unaffected (see Table \ref{tab_parameters_grid} and Figures \ref{figBolMod} and \ref{figVelMod}). In particular, $L_M$ is an increasing function of $M_{Ni}$, growing by a factor of $\sim 20$ when passing from $M_{Ni}= 10^{-4}$\Msun to $M_{Ni}= 0.5$\msun. However, as expected, the value of $L_M$ remains essentially unchanged when passing from $M_{Ni}= 10^{-4}$\Msun to $M_{Ni}= 10^{-2}$\,\msun, and it increases only by a factor of $\sim 2.2$ for $M_{Ni}$ lying between $10^{-4}$\,\Msun to $0.04$\,\msun, showing that the heating effects due to the $^{56}$Ni are noticeable only for sufficiently high amount of $^{56}$Ni. In our models with $M_{ej}$, $E$, and $R$, respectively, fixed to $16$\,\msun, $1$\,foe, and $3$x$10^{12}$\,cm, this ``threshold'' value of $^{56}$Ni is a few hundredths of solar masses. However, for different values of $M_{ej}$ and/or $E$ and/or $R$, its value may change because the impact of the heating effects due to the $^{56}$Ni on the total energetic budget of the ejected material can be different. The behaviour of $t_M$ is very similar to that of $L_M$, but $t_M$ grows by only a factor of $\sim 4$ when passing from $M_{Ni}= 10^{-4}$\,\Msun to $M_{Ni}= 0.5$\,\msun. Instead $v_M$ does not display a monotonic trend with the $^{56}$Ni mass, but it seems to suddenly decrease by a factor $\sim 1.5$ as soon as the heating effects due to the $^{56}$Ni are noticeable (i.e.~for $^{56}$Ni amount greater than the above mentioned ``threshold'' value). As a consequence, among all the derived scaling relations, those depending on $L_M$ and/or depending on $t_M$ and $v_M$ with larger exponents are affected to a greater extent in neglecting the heating effects due to the $^{56}$Ni in the analytical models and, consequently, are less accurate. This agrees with our above findings, according to which the first two relations in set (\ref{setArnett}) and the third relation of set (\ref{setPopov}) appear the best in terms of accuracy.\par

Nonetheless the deviation between the values of $E$, $M_{ej}$, and $R$ derived from the scaling relations and those estimated with hydrodynamical modelling, tends to decrease when the reference SN has an amount of $^{56}$Ni similar to that of the event for which the scaling relations are used to derive the triplets $E$, $M_{ej}$, and $R$ (see Figures \ref{figratio87A} to \ref{figratioOGLE073}). In particular, Figures \ref{figratio87A} (black diamonds) and \ref{figratio09E} show that the ratio between the values of the triplets $E$, $M_{ej}$, and $R$ derived from the scaling relations of set (\ref{setArnett}) and those estimated with hydrodynamical modelling is about 1 (namely within the range 0.4-2.5 at the most) when the ratio between the amount of $^{56}$Ni of the reference SN and that of the event for which the scaling relations are used to derive the parameters $E$, $M_{ej}$, and $R$, is also near to 1 (namely in the range $\sim$ 0.4-1.3). The only exception seems to be the $R$ parameter for OGLE073, given that the ratio between the value of $R$ derived from the scaling relations of set (\ref{setArnett}) and that estimated with hydrodynamical modelling is very different from 1 (namely $\sim$ 0.02-0.03) when the reference SN has an amount of $^{56}$Ni similar to that of OGLE073 (see Figure \ref{figratioOGLE073}). This behavior is probably related to the peculiarity of this SN, that is a ``non-conventional'', highly massive, high-energy event (cf.~Section \ref{intro}). As a consequence, in order to retrieve a sufficiently accurate value of $R$ when using the third scaling relation of set (\ref{setArnett}), it is important to use a reference SN with not only a similar amount of $^{56}$Ni but also with values of $E$, $M_{ej}$, and $R$ that are nearer in the parameter space to those describing OGLE073. In practice, this implies to adopt a reference SN with a bolometric light curve similar to that of OGLE073 in terms of both shape and luminosity at the epoch of the bolometric light-curve maximum. For example, using the model having the bolometric light curve reported in Figure \ref{figmodelloNew} as reference SN for OGLE073, the value of $R$ derived from the scaling relation of set (\ref{setArnett}) is equal to $39.0(\pm 6.7)$x$10^{12}$ cm, fully in agreement with the estimate of $38.0(^{+8.0}_{-10.0})$x$10^{12}$ cm inferred through procedures of hydrodynamical modelling (cf.~Section \ref{observations} and Table \ref{tab_parameters_modelling}). Thus, it seems to be possible to use all the three scaling relations of set (\ref{setArnett}) to simultaneously retrieve sufficiently accurate values of $E$, $M_{ej}$, and $R$, assuming that the reference SN is conveniently chosen. As already noticed (cf.~Section \ref{modelling}), this could be very useful to characterize long-rising SNe for which the spectro-photometric behavior is well known only at the epoch of the bolometric light-curve maximum. Similar considerations are also valid for the scaling relations of sets (\ref{setPopov}) and (\ref{setPopov_phot}) but, in order to retrieve sufficiently accurate values of $M_{ej}$, $R$ and --- even more --- $E$, the ratio between the amount of $^{56}$Ni of the reference SN and that of the event at which the scaling relations are applyed, has to be very close to 1 (namely in the range $\sim$ 0.4-1.1 at the most; see red cirles and green squares in Figure \ref{figratio87A}). Thus, also for the sets (\ref{setPopov}) and (\ref{setPopov_phot}), it seems to be possible to use the three scaling relations of each set to simultaneously retrieve sufficiently accurate values of $E$, $M_{ej}$, and $R$, provided that the reference SN is conveniently chosen. We remind (cf.~Section \ref{modelling}) that the usage of scaling relations of set (\ref{setPopov_phot}) could be very useful to characterize long-rising SNe for which only the photometric behavior is well known. Indeed the development of a method for deriving $E$, $M_{ej}$, and $R$ based solely on photometric data could be of primary importance in the context of future SNe surveys, that potentially follow the photometric evolution of thousands or more SNe with a limited (or without) spectroscopic follow-up.\par

\subsection{Comparative analysis}
\label{systematics}

After having analysed the robustness of the scaling relations of sets (\ref{setArnett}), (\ref{setPopov}), and (\ref{setPopov_phot}) in Section \ref{scaling_rel}, we use the values of $E$, $M_{ej}$, and $R$ inferred applying the most accurate and precise relations to our sample of SN 1987A-like objects. In particular, we consider the first two relations of set (\ref{setArnett}) and the third relation of set (\ref{setPopov}), using SN 1987A as reference. For SNe 1987A, 2009E, and OGLE073, we consider the values of $E$, $M_{ej}$, and $R$ already estimated through our hydrodynamical modelling (cf.~Section \ref{observations}).\par

\begin{figure}
 \includegraphics[angle=-90,width=85mm]{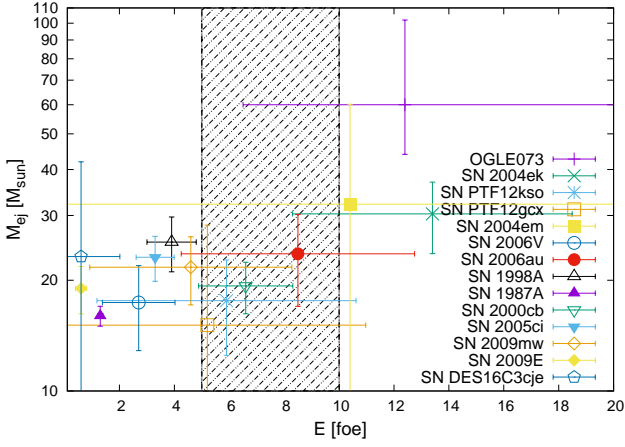} 
 \caption{$M_{ej}$ versus $E$ for the sample of SN 1987A-like objects considered in this work. The dashed area constrains the $5 \leq E/[foe] \leq 10$ range (see text for details).
 \label{figEMplane}}
\end{figure}

\begin{figure}
 \includegraphics[angle=-90,width=85mm]{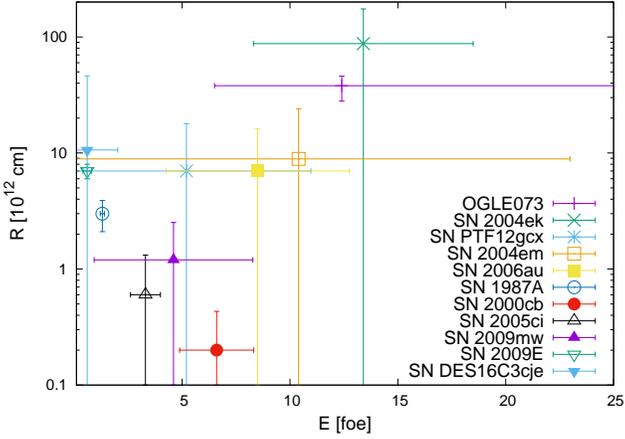} 
 \caption{As for Figure \ref{figEMplane}, but for $R$ versus $E$. SNe PTF12kso, 2006V and 1998A are not reported because $R$ can not be determined using the third relation of set (\ref{setPopov}), given that $L_m$ is not known (cf.~Section \ref{scaling_rel}). For SNe 2000cb, 2005ci and 2009mw the reported value of $R$ is an upper limit (cf.~Tables \ref{tab_observations} and \ref{tab_Popov}).
 \label{figERplane}}
\end{figure}

\begin{figure}
 \includegraphics[angle=-90,width=85mm]{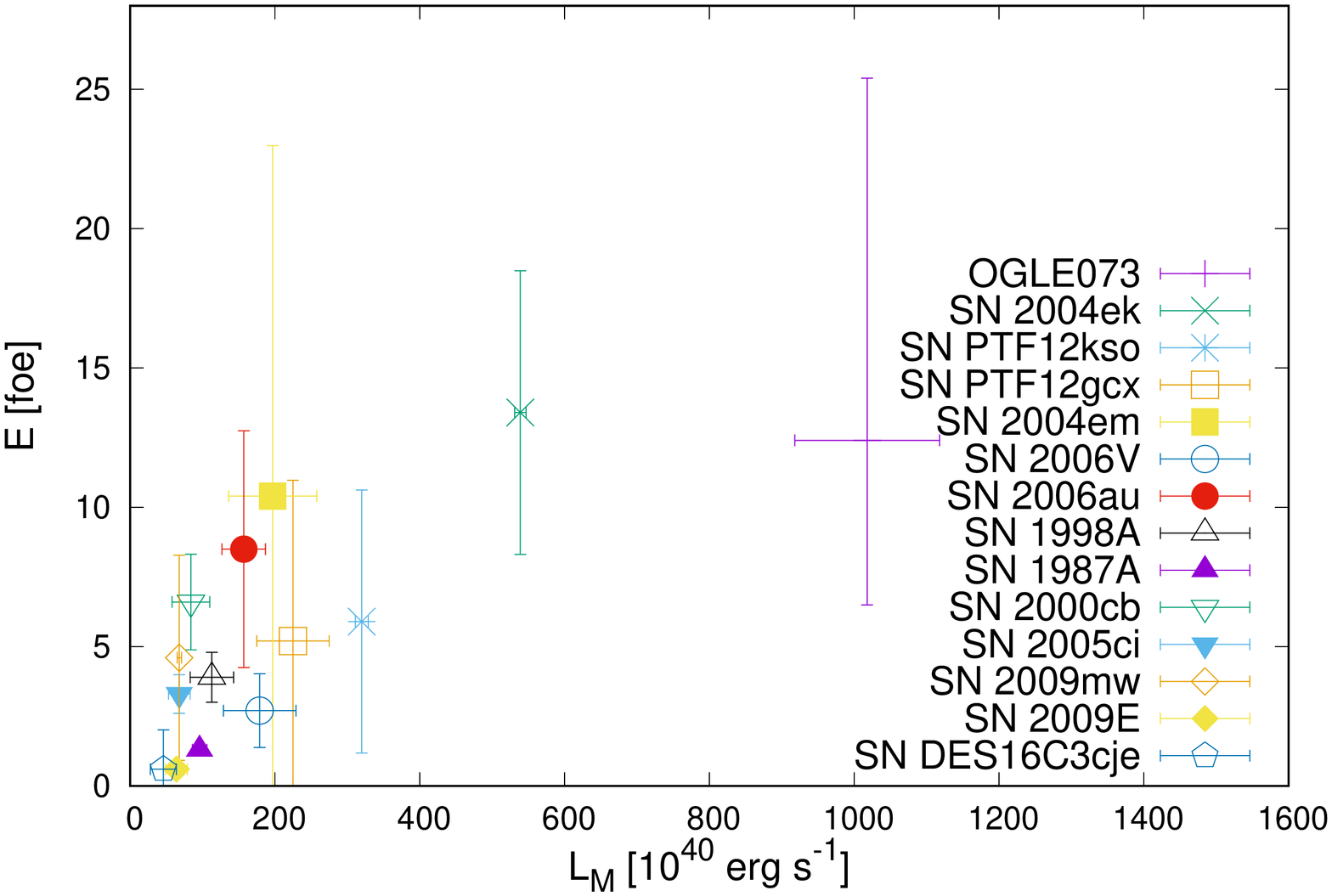} 
 \caption{As for Figure \ref{figEMplane}, but for $E$ versus $L_M$ (see text for details).
 \label{figELplane}}
\end{figure}

\begin{figure}
 \includegraphics[angle=-90,width=85mm]{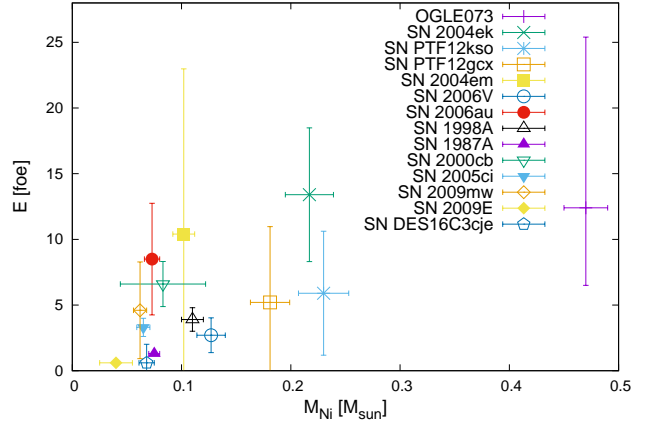} 
 \caption{As for Figure \ref{figEMplane}, but for $E$ versus $\mathcal{M}_{Ni}$ (see text for details).
 \label{figENplane}}
\end{figure}

The data reported in Figures \ref{figEMplane} and \ref{figERplane} indicate that SN 1987A-like objects have parameters at explosion covering a wide range of values, as found for other sub-classes of H-rich SNe like the Type II plateau SNe \citep[see e.g.][]{spiro14,pumo17}. In particular, the long-rising SNe of our sample are placed in the $M_{ej}$-$E$ plane along a diagonal band in an almost continuous distribution, moving from low-energetic ($\sim 0.5$-$0.6$\,foe) SNe with realtively low-mass ejecta ($\sim 15$-$25$\,\msun) to high-massive ($\gtrsim 30$\,\msun), high-energy ($\gtrsim$ 10\,foe) events. With the warning that our sample could be too small to draw final conclusions, SN 1987A-like objects form a ``family'' of explosive events where the main parameter ``guiding'' the distribution seems to be the explosion energy $E$. A correlation between $E$ and the observed quantities such as $L_M$ and the amount of $^{56}$Ni present in the SN ejecta, $\mathcal{M}_{Ni}$, is quite evident (see Figures \ref{figELplane} and \ref{figENplane}). Indeed, both quantities tend to increase when increasing $E$ and, from a statistical point of view, the correlations $E$-$L_M$ and $E$-$\mathcal{M}_{Ni}$ are respectively significant at 99 and 95 per cent confidence level (the null hypothesis two-tailed probability inferred from the Pearson correlation coefficient are respectively $\simeq 0.003 < 0.01$ and $\simeq 0.014 < 0.05$). Roughly speaking, it is possible to identify three subgroups of events according to the $E$ value. The first one is formed by substantial clones of SN 1987A, that can be explained in terms of neutrino-driven core-collapse explosion with $E$ always ranging from several tenths of foe up to some foe. The second subgroups is formed by the tail of high-energy ($\gtrsim$ 10\,foe) events, whose physical properties of the progenitor at explosion (primary the explosion energy and the ejected mass) are difficult to explain within the neutrino-driven core-collapse paradigm \citep[see also][and references therein]{terreran16}. In particular, for this subgroup of events, the explosion energies are a factor $\sim$ 3-6 higher than the maximum value expected in canonical neutrino-driven core-collapse explosions. Moreover, according to the current state-of-the-art evolutionary theory, their progenitors should explode as H-free SNe after non-negligible mass-loss during their pre-SN evolution, so it is still puzzling how they can retain a sufficiently large fraction of their initial (i.e.~at the star birth) H-rich outer stellar layers. The third subgroup is formed by ``transitional'' events with $E$ in the range $\sim$ 5-10 foe (see the dashed area in Figure \ref{figEMplane}), that essentially bridge the standard SN 1987A-like objects with the tail of high-energy events. For SNe of this subgroup, the uncertainties on the values of $E$ do not allow us to firmly establish whether they can be explained in terms of conventional neutrino-driven core-collapse events or not. Data in Figure \ref{figERplane} also show that the high-energy events are always linked to particularly extended progenitors having $R \sim 10^{13}$-$10^{14}$\,cm. Considering that these high-energy SN 1987A-like objects are also Ni-rich (see Table \ref{tab_parameters_modelling}), our findings agree with \citet[][]{taddia16} according to which long-rising SNe can also arise from progenitors with very extended radii (of the order of thousands of \rsun) when a sufficiently large amount of \chem{56}{Ni} ($\gtrsim 0.1$-$0.2$\,\Msun) is synthesized in the explosion.\par


\begin{figure}
 \includegraphics[angle=-90,width=85mm]{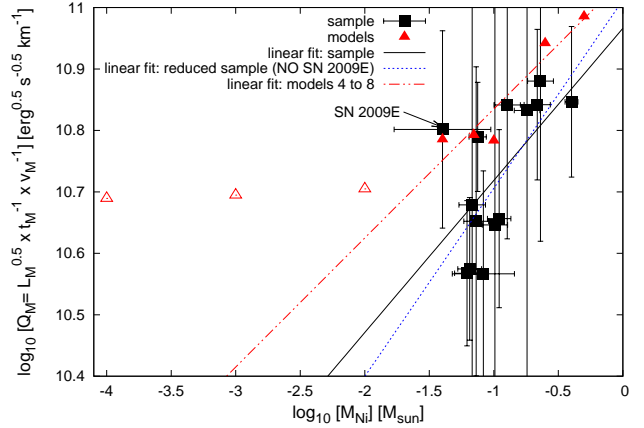}
 \caption{Correlation between $\mathcal{M}_{Ni}$ and $Q_M$ (see the text for details) both for the sample of SN 1987A-like objects considered in this work (black squares) and for the models 4 to 8 of Table~\ref{tabmodels} (red filled triangols). The best linear fits mentioned in the text are also shown with lines of different styles. Moreover, the position of the ``Ni-poor'' models (i.e.~models 1 to 3 of Table~\ref{tabmodels}) is reported in the log$_{10}$($\mathcal{M}_{Ni}$)-log$_{10}$($Q_M$) plane for comparison purpose (red open triangols).
 \label{figNi}}
\end{figure}

Furthermore, in the sample of SN 1987A-like objects considered in this work, we note (see black squares in Figure \ref{figNi}) a correlation between $\mathcal{M}_{Ni}$ and the physical quantity $Q_M\equiv L_M^{1/2} t_M^{-1} v_M^{-1}$, which is a linear combination of $L_M$, $t_M$ and $v_M$. As such, $Q_M$ depends only on the spectro-photometric characteristics of the SN at the epoch of the light-curve maximum and, from a physical point of view, it can be directly correlated to the Poynting vector's modulus. Indeed, $Q_M$ is the square root of luminous power on a surface, being the square root of the ratio between the SN luminosity at the epoch of the light-curve maximum and the square of its photospheric radius $R_{ph}(t_M)=v_M t_M$ at the same epoch. From a statistical point of view, the correlation between $\mathcal{M}_{Ni}$ and $Q_M$ is significant at 95 per cent confidence level (the null hypothesis two-tailed probability inferred from the Pearson correlation coefficient is $\simeq 0.019$) and the best linear fit describing the relation (black solid line in Figure \ref{figNi}) is given by the equation

\begin{equation}
\label{eq:Nisample}
\log_{10}\mathrm{Q}_{M} = (0.25 \pm 0.09) \log_{10}\mathcal{M}_{Ni} + (10.97 \pm 0.10)\, .
\end{equation}

However, the data in Figure \ref{figNi} show a not negligible scatter with a root-mean-square (rms) deviation around the fit $\simeq 0.096$. Although $Q_M$ tends to increase with $\mathcal{M}_{Ni}$ in most cases, there are some exceptions. The most important one is SN 2009E, which is characterised by a residual greater than the error on $Q_M$. Excluding SN 2009E from the sample, the $\log_{10}\mathrm{Q}_{M}$-$\log_{10}\mathcal{M}_{Ni}$ correlation becames significant at 99 per cent confidence level (the null hypothesis two-tailed probability inferred from the Pearson correlation coefficient is $\simeq 0.001$) and the best linear fit (blue dotted line in Figure \ref{figNi}) changes into the equation

\begin{equation}
\label{eq:NisampleNo09E}
\log_{10}\mathrm{Q}_{M} = (0.31 \pm 0.09) \log_{10}\mathcal{M}_{Ni} + (11.01 \pm 0.09)\, .
\end{equation}

\noindent In this last case, the rms deviation of the data around the fit is $\simeq 0.076$ and decreases compared to that obtained for the previous relation (\ref{eq:Nisample}). The parameters of the best fit also change and the fractional error on the slope decreases from $\sim$ 39 to $\sim$ 29 per cent. Nevertheless, relations (\ref{eq:Nisample}) and (\ref{eq:NisampleNo09E}) are statistically mutually consistent.

The $\log_{10}\mathrm{Q}_{M}$-$\log_{10}\mathcal{M}_{Ni}$ correlation, when inverted, may represent an interesting tool for estimating the amount of $^{56}$Ni in the ejecta of long-rising SNe whitout the need of having information on the tail luminosity. However, the change in the slope when considering the whole sample [relation (\ref{eq:Nisample})] or excluding SN 2009E [relation (\ref{eq:NisampleNo09E})], is not negligible, although statistically not major. This suggests that the slope of the above reported correlation could be sensitive to the outliers, probably because of the still numerically limited sample. For all the above reasons, the $\log_{10}\mathrm{Q}_{M}$-$\log_{10}\mathcal{M}_{Ni}$ correlation should not be considered as a ``ready-to-use recipe'' for deriving accurate information about the amount of $^{56}$Ni in the SN ejecta. Its usage should be done cum grano salis, and it is desiderable to further test the relation against a larger sample of well-observed events.

Also our radiation-hydrodynamical models show the $\log_{10}\mathrm{Q}_{M}$-$\log_{10}\mathcal{M}_{Ni}$ correlation when the heating effects due to the $^{56}$Ni are not negligible (i.e.~for models 4 to 8 of Table~\ref{tabmodels}, cf.~also Section \ref{scaling_rel}). In this case the best linear fit (red dash-dotted line in Figure \ref{figNi}) is

\begin{equation}
\label{eq:Nimodel1418}
\log_{10}\mathrm{Q}_{M} = (0.21 \pm 0.04) \log_{10}\mathcal{M}_{Ni} + (11.04 \pm 0.04)\, ,
\end{equation}

\noindent while the rms deviation of the models around the fit is $\simeq 0.038$, and the fractional error on the slope is $\sim$ 21 per cent. The parameters of the best fit relation (\ref{eq:Nimodel1418}) change with respect to those obtained when considering the sample of real SN 1987A-like objects [i.e.~parameters of relations (\ref{eq:Nisample}) and (\ref{eq:NisampleNo09E})], but the relations (\ref{eq:Nimodel1418}), (\ref{eq:Nisample}) and (\ref{eq:NisampleNo09E}) are statistically mutually consistent. However, although statistically not significant, the change in the slope when considering the radiation-hydrodynamical models is not negligible, showing that the slope's value in relation (\ref{eq:Nimodel1418}) could be somewhat dependent on the explored model parameter space.\par

Last but not least, the ``Ni-poor'' models (i.e.~models 1 to 3 of Table~\ref{tabmodels}) do not show the $\log_{10}\mathrm{Q}_{M}$-$\log_{10}\mathcal{M}_{Ni}$ correlation and are characterized by having an almost constant $\mathrm{Q}_{M}$ value (red open triangles in Figure \ref{figNi}), as expected when the amount of $^{56}$Ni is too low for noticeably affecting the total energetic budget of the ejecta (cf.~Section \ref{scaling_rel}) and consistent with what predicted by the relation (\ref{eqForNi}), based on the analytic model of \citet{popov93} that does not consider the heating effects linked to the $^{56}$Ni decay (cf.~Section \ref{modelling}). All of this also suggests that new analytic models including appropriately \chem{56}{Ni} decay heating effects (see Section \ref{summary} for further details), should be developed for shedding more light on the physical origin of the $\log_{10}\mathrm{Q}_{M}$-$\log_{10}\mathcal{M}_{Ni}$ relation.

\section{Summary and further comments}
\label{summary}

In order to improve our knowledge about long-rising SNe resembling SN 1987A, we conduct a comparative study, using the best scaling relations in terms of accuracy and precision to infer the SN progenitor’s physical properties at the explosion (namely the ejected mass $M_{ej}$, the progenitor radius at the explosion $R$ and the total explosion energy $E$) for SN 1987A-like objects.\par

To select such best relations, we first derive and test different scaling relations based on the analytic models describing the post-explosive evolution of H-rich SN ejecta of \citet{arnett80} and \citet{popov93}. The main findings can be summarized as it follows.\par

(a.) It is possible to derive three triplets --- one based on the model of \citet{arnett80} and two based on that of \citet{popov93} --- of scaling relations, for a total of nine indipendent and interchangeable relations, most of which are new. They are useful to simultaneously retrieve the values of $E$, $M_{ej}$, and $R$ for a long-rising SN, provided that these three values are independently known at least for another long-rising object, referred to as reference SN.\par

(b.) The robustness and feasibility of these sets of scaling relations are different and depend on various factors as neglecting heating effects linked to the presence of \chem{56}{Ni} when modelling the ejecta evolution. In particular, the set based on the model of \citet{arnett80} [see relationships (\ref{setArnett})] has the clear advantage that it can be used once the spectro-photometric behavior of the long-rising SN is known only at the epoch of the bolometric light-curve maximum, but the relationship for $R$ is sufficiently accurate only if the reference SN is conveniently chosen. On the contrary, in the first set based on the model of \citet{popov93} [see relationships (\ref{setPopov})], the relationship for $R$ is more accurate but, in order to use this relation, it is necessary to well sample the bolometric light-curve also a long time before the maximum. The second set based on the model of \citet{popov93} [see relationships (\ref{setPopov_phot})] has instead the clear advantage that can be used once only the photometric behavior of the long-rising SN is known. However, the relationships for $E$, $M_{ej}$, and $R$, are  sufficiently accurate only if the reference SN is conveniently chosen.\par

(c.) Globally, among the nine relations, the best ones in terms of accuracy are the relationships for $E$ and $M_{ej}$ based on the model of \citet{arnett80}, and that for $R$ in the first set of scaling relations based on the model of \citet{popov93}.\par

After individuating the best scaling relations in terms of accuracy, we apply them to a selected sample of SNe resembling SN 1987A, enabling us to conduct a comparative study. The main findings can be summarized as it follows.\par

(d.) SN 1987A-like objects have parameters at explosion covering a wide range of values ($E \sim 0.5$-$15$ foe, $R \sim 0.2$-$100 \times 10^{12}$ cm, and $M_{ej} \sim 15$-$55$\msun), as found for other sub-classes of SNe.\par

(e.) The main parameter ``guiding'' their distribution seems to be $E$.\par 

(f.) There is a high-massive ($\gtrsim 30$\,\Msun), high-energy ($\gtrsim 10$\,foe) tail of events, always linked to extended progenitors with radii at explosion $\sim 10^{13}$-$10^{14}$\,cm, that challenge standard theories of neutrino-driven core-collapse and stellar evolution.\par

In the sample of SN 1987A-like objects considered in this work, we also find a correlation between the amount of $^{56}$Ni in the SN ejecta and the spectrophotometric features of the SN at the epoch of the light-curve maximum, that may represent an interesting tool for estimating the amount of $^{56}$Ni whitout having information on the luminosity of SN 1987A-like objects in the radioactive tail.\par

Although the sample of SN 1987A-like objects is one of the biggest and most complete ever considered in literature, it could be still too small to draw final conclusions. For this reason, other future studies based on larger samples of long-rising SNe resembling SN 1987A, are needed to confirm our results. Moreover, it should be useful to further check our results deriving the values of $E$, $M_{ej}$, and $R$ through more precise and accurate approaches like the ``homogeneous and self-consistent'' hydrodynamical modelling. In other words, it should be useful to apply the same hydrodynamical modelling to the whole sample of SN 1987A-like objects, using numerical simulations that include the SN explosion and the explosive nucleosynthesis, starting from pre-SN models evaluated through stellar evolution codes (see \citealt[][]{PZ11}, \citeyear[][]{PZ12} and \citealt[][]{pumo17}, for further details). Furthermore, for a better understanding of the physical origin of the correlation between the amount of $^{56}$Ni in the SN ejecta and the spectrophotometric features of the SN at maximum, it would be desirable to develop analytic models including the heating effects due to the \chem{56}{Ni} decay on the SN ejecta evolution during the whole post-explosive phase.  
  
\section*{Acknowledgments}
We are grateful to the {\it Laboratori Nazionali del Sud - Istituto Nazionale di Fisica Nucleare} for the use of HPC facilities. We also thank Francesco Taddia for provided published data in electronic format of various SN 1987A-like objects. M.L.P.~acknowledges support from the plan ``{\it programma ricerca di ateneo UNICT linea 2 PIA.CE.RI. 2020-2022}'' of the Catania University (project ASTRI, P.I. F.~Leone, ID 55722062158).

\section*{Data availability}
The data underlying this article are available in the article.

\bibliographystyle{aa}



\appendix
\section{Extra material on the scaling relations}
\label{app_formulas}

To derive Eq.~(\ref{eqForNi}), it is sufficient to consider that set (\ref{setPpovdegenerate}) can be also written as

\begin{equation}
\label{setPpovdeg_equal_symbol}
\left\{
\begin{array}{l}
\displaystyle
(a)\> \> E^3 M_{ej}^{-5} = k_a v_M^4 t_M^{-4}\\
(b)\> \> R^2 M_{ej} E    = k_b L_M^{3} v_M^{-2} t_M^2\\
(c)\> \> R^3 M_{ej}^4    = k_c v_M^4 t_M^{14}
\end{array}
\right.
\end{equation}
where $k_a$, $k_b$, and $k_c$ are numerical constants different from zero. Taking the logarithm in both sides of each relation, set (\ref{setPpovdeg_equal_symbol}) can be easily converted into the following linear system:
\begin{equation}
\label{systemPpovdeg}
\left\{
\begin{array}{l}
\displaystyle
(a')\> \> 3y - 5z = A\\
(b')\> \> 2x + y  + z = B\\
(c')\> \> 3x + 4z = C 
\end{array}
\right.
\end{equation}
where $x=log(R)$, $y=log(E)$ and $z=log(M_{ej})$ are variables, while $A=log(v_M^4 t_M^{-4})+log(k_a)$, $B=log(L_M^3 v_M^{-2} t_M^2)+log(k_b)$ and $C=log(v_M^4 t_M^{14})+log(k_c)$ are the constant terms, because they depend only on $L_M$, $v_M$ and $t_M$ (that are quantities fixed from the observational data) and $k_a$, $k_b$ and $k_c$ (that are numerical constants). 
The system (\ref{systemPpovdeg}) can be solved with the Gaussian elimination method \citep[see e.g.][for details]{press96}. Combining the equations as $(a')-3(b')+2(c')\> \rightarrow \> (3y-5z)-3(2x+y+z)+2(3x+4z)=A-3B+2C\>  \rightarrow\> 0=A-3B+2C$, it can be either impossible (i.e., it does not admit solutions) if $A-3B+2C\neq0$, or degererate (i.e., it admits infinite solutions) if $A-3B+2C=0$. The case presented in this paper coincides with the last one, being rappresentative of a real physical case that must admit solutions. As a consequence, the following series of relations are also valid:
$(a')-3(b')+2(c')=0  \rightarrow exp[(a')-3(b')+2(c')]=1 \rightarrow 
(a) (b)^{-3} (c)^2= k_a^{-1} k_b^{3}  k_c^{-2}$, with the right-side term  $k_a^{-1} k_b^{3}  k_c^{-2}$ being equal to a numerical constant different from zero $\rightarrow (a) \propto (b)^3 (c)^{-2} \equiv$ Eq.~(\ref{eqForNi}); Q.E.D.

From a physical point of view, the relations of set (\ref{setPpovdeg_equal_symbol}) are based on a model where the SN ejecta are considered to emit as a black-body (cf.~Section \ref{modelling} and see also \citealt{popov93}). The validity of the relation $A-3B+2C=0$ and, consequently, the degeneration of set (\ref{setPpovdeg_equal_symbol}), are consistent with the black-body hypothesis. In particular, the relation $A-3B+2C=0$ also implies the validity of the $L_M\propto v_M^2 t_M^2$ relation, where the SN luminosity at maximum $L_M$ is proportional to the square of photospheric radius at the same time (equal to the term $v_M t_M$), which is consistent with the Stefan-Boltzmann law. Indeed the following chain of relations is valid: $A-3B+2C=0 \rightarrow log\left(v_M^{18} t_M^{18} L_M^{-9} \right) + log\left(k_a k_b^{-3} k_c^2\right) =0 \rightarrow 
log\left(v_M^{2} t_M^{2} L_M^{-1} \right) = log\left(k_a k_b^{-3} k_c^2\right)^{-1/9}$ with the right-side term $log\left(k_a k_b^{-3} k_c^2\right)^{-1/9}$ being equal to a numerical constant different from zero $\rightarrow L_M \propto v_M^2 t_M^2$; Q.E.D.  

The $L_M\propto v_M^2 t_M^2$ relation, equivalent to $v_M \propto L_M^{1/2} t_M^{-1}$, is the same used to derive the scaling relations of set (\ref{setPopov_phot}) based only on the photometric behavior of the long-rising SN (cf.~Section \ref{modelling}). Last but not least, we note that it can be easly derived by also applying Eq.~(\ref{eqForNi}) directly to the right-side terms of the relations (a), (b) and (c) in set (\ref{setPpovdegenerate}). One hence obtains the following series of relations: $v_M^{4} t_M^{-4} \propto \left(L_M^{3} v_M^{-2} t_M^{2}\right)^{3} \left(v_M^{4} t_M^{14}\right)^{-2} \rightarrow L_M\propto v_M^2 t_M^2  \rightarrow v_M \propto L_M^{1/2} t_M^{-1}$; Q.E.D.

\section{Extra material on the SN progenitor's physical properties inferred through scaling relations} 

\label{app_scaling_rel}

In this appendix we present additional material on the values of $E$, $M_{ej}$, and $R$ inferred by means of the relations of sets (\ref{setArnett}), (\ref{setPopov}) and (\ref{setPopov_phot}) --- hereafter indicated as Arnett set, Popov set, and ``pure photometric'' Popov set, respectively --- adopting the radiation-hydrodynamical models of Section \ref{models} as reference SNe. In particular, in Tables \ref{tab_mod_Arnett_lowNi} to \ref{tab_mod_PurePhot_Popov_highNi} we report the results obtained for the sample of well-observed SN 1987A-like objects considered in this work (cf.~Section \ref{observations}). In Tables \ref{tab_grid_mod_Arnett} to \ref{tab_grid_mod_PurePhot_Popov} we report the results obtained for the grid of radiation-hydrodynamical models (i.e.~applying the scaling relations to the simulated bolometric luminosities and velocities).\par


\begin{table*}
   \centering
   \caption{Same as Table~\ref{tab_Arnett}, but for values of $E$, $M_{ej}$, and $R$ obtained with the Arnett set considering the ``low-Ni'' models (i.e.~models 1 and 2 of Table~\ref{tabmodels} having $M_{Ni}<0.01$\msun) as reference SNe. The error percentages (reported in the last three columns) do not depend on the choice of the reference model but only on the set of scaling equations and the SN measurement errors (see Table~\ref{tab_observations}).}
   \begin{tabular}{lcccccccrrrr}
   \hline\hline
   
   SN & \multicolumn{1}{c}{$E$} & \multicolumn{1}{c}{$M_{ej}$} & \multicolumn{1}{c}{$R$} & \multicolumn{1}{c}{$E$} & \multicolumn{1}{c}{$M_{ej}$} & \multicolumn{1}{c}{$R$} & & \multicolumn{1}{c}{$\pm E$} & \multicolumn{1}{c}{$\pm M_{ej}$} & \multicolumn{1}{c}{$\pm R$}  \\
      & \multicolumn{1}{c}{[foe]} & \multicolumn{1}{c}{[\msun]} & \multicolumn{1}{c}{[$10^{12}cm$]} & \multicolumn{1}{c}{[foe]} & \multicolumn{1}{c}{[\msun]} & \multicolumn{1}{c}{[$10^{12}cm$]} & & \multicolumn{1}{c}{[$\%$]} & \multicolumn{1}{c}{[$\%$]} & \multicolumn{1}{c}{[$\%$]}\\    
     
   \hline
   & \multicolumn{3}{c}{ref.~SN: Model 1} & \multicolumn{3}{c}{ref.~SN: Model 2} & & \multicolumn{3}{c}{Error percentages} \\
   \hline
   OGLE073     & $50.1$ & $194.1$ & $36.9$ & $50.1$ & $184.7$ & $32.2$ & & $27$  & $19$ & $17$  \\
   2004ek     & $25.2$ & $111.2$ & $22.2$ & $25.2$ & $105.8$ & $19.4$ & & $32$  & $17$ & $19$  \\
   PTF12kso   & $11.0$ & $64.5$  & $17.5$ & $11.0$ & $61.4$  & $15.3$ & & $78$  & $26$ & $52$  \\
   PTF12gcx   & $9.8$  & $55.5$  & $12.0$ & $9.8$  & $52.8$  & $10.4$ & & $109$ & $87$ & $52$  \\
   2004em     & $19.5$ & $118.2$ & $11.2$ & $19.5$ & $112.4$ & $9.8$  & & $119$ & $86$ & $67$  \\
   2006V      & $5.6$  & $65.8$  & $19.8$ & $5.6$  & $62.6$  & $17.2$ & & $44$  & $22$ & $39$  \\
   2006au     & $17.7$ & $89.4$  & $7.4$  & $17.7$ & $85.1$  & $6.5$  & & $67$  & $24$ & $48$  \\
   1998A      & $27.2$ & $144.1$ & $5.6$  & $27.2$ & $137.1$ & $4.9$  & & $12$  & $11$ & $27$  \\
   1987A      & $2.4$  & $58.8$  & $21.5$ & $2.4$  & $55.9$  & $18.8$ & & $18$  & $12$ & $14$  \\
   2000cb     & $12.4$ & $71.1$  & $4.5$  & $12.4$ & $67.6$  & $3.9$  & & $18$  & $9$  & $33$  \\
   2005ci     & $6.2$  & $85.0$  & $8.8$  & $6.2$  & $80.8$  & $7.7$  & & $10$  & $5$  & $23$  \\
   2009mw     & $7.0$  & $79.8$  & $7.3$  & $7.0$  & $75.9$  & $6.3$  & & $32$  & $16$ & $21$  \\
   2009E      & $1.2$  & $50.7$  & $25.9$ & $1.2$  & $48.2$  & $22.5$ & & $32$  & $13$ & $31$  \\
   DES16C3cje & $1.1$  & $85.1$  & $33.5$ & $1.1$  & $81.0$  & $29.2$ & & $235$ & $80$ & $161$ \\

   \hline                                   
 \end{tabular}
 \label{tab_mod_Arnett_lowNi}
\end{table*}

\begin{table*}
   \centering
   \caption{Same as Table~\ref{tab_mod_Arnett_lowNi}, but for the models with $M_{Ni} \geq 0.01$\Msun (i.e.~models 3 to 8 of Table~\ref{tabmodels}) as reference SNe. See Table~\ref{tab_mod_Arnett_lowNi} for the errors on the values of $E$, $M_{ej}$, and $R$.}
   \begin{tabular}{lccccccccc}
   \hline\hline
   SN & \multicolumn{1}{c}{$E$} & \multicolumn{1}{c}{$M_{ej}$} & \multicolumn{1}{c}{$R$} & \multicolumn{1}{c}{$E$} & \multicolumn{1}{c}{$M_{ej}$} & \multicolumn{1}{c}{$R$} & \multicolumn{1}{c}{$E$} & \multicolumn{1}{c}{$M_{ej}$} & \multicolumn{1}{c}{$R$} \\
      & \multicolumn{1}{c}{[foe]} & \multicolumn{1}{c}{[\msun]} & \multicolumn{1}{c}{[$10^{12}cm$]} & \multicolumn{1}{c}{[foe]} & \multicolumn{1}{c}{[\msun]} & \multicolumn{1}{c}{[$10^{12}cm$]} & \multicolumn{1}{c}{[foe]} & \multicolumn{1}{c}{[\msun]} & \multicolumn{1}{c}{[$10^{12}cm$]} \\
   \hline   
    & \multicolumn{3}{c}{ref.~SN: Model 3} & \multicolumn{3}{c}{ref.~SN: Model 4} & \multicolumn{3}{c}{ref.~SN: Model 5} \\    
   \hline
   OGLE073     & $46.7$ & $171.8$ & $29.7$ & $66.1$ & $74.5$  & $4.9$  & $45.3$ & $54.8$  & $3.6$ \\
   2004ek     & $23.5$ & $98.4$  & $17.9$ & $33.3$ & $42.7$  & $3.0$  & $22.8$ & $31.4$  & $2.2$ \\
   PTF12kso   & $10.3$ & $57.1$  & $14.1$ & $14.6$ & $24.8$  & $2.3$  & $10.0$ & $18.2$  & $1.7$ \\
   PTF12gcx   & $9.1$  & $49.1$  & $9.6$  & $12.9$ & $21.3$  & $1.6$  & $8.8$  & $15.7$  & $1.2$ \\
   2004em     & $18.2$ & $104.6$ & $9.0$  & $25.7$ & $45.4$  & $1.5$  & $17.6$ & $33.4$  & $1.1$ \\
   2006V      & $5.2$  & $58.2$  & $15.9$ & $7.3$  & $25.3$  & $2.6$  & $5.0$  & $18.6$  & $1.9$ \\
   2006au     & $16.5$ & $79.1$  & $6.0$  & $23.3$ & $34.3$  & $1.0$  & $16.0$ & $25.2$  & $0.7$ \\
   1998A      & $25.4$ & $127.6$ & $4.5$  & $35.9$ & $55.3$  & $0.7$  & $24.6$ & $40.7$  & $0.5$ \\
   1987A      & $2.3$  & $52.0$  & $17.3$ & $3.2$  & $22.6$  & $2.9$  & $2.2$  & $16.6$  & $2.1$ \\
   2000cb     & $11.5$ & $62.9$  & $3.6$  & $16.3$ & $27.3$  & $0.6$  & $11.2$ & $20.1$  & $0.4$ \\
   2005ci     & $5.7$  & $75.2$  & $7.1$  & $8.1$  & $32.6$  & $1.2$  & $5.6$  & $24.0$  & $0.9$ \\
   2009mw     & $6.5$  & $70.6$  & $5.8$  & $9.2$  & $30.6$  & $1.0$  & $6.3$  & $22.5$  & $0.7$ \\
   2009E      & $1.1$  & $44.8$  & $20.8$ & $1.5$  & $19.5$  & $3.4$  & $1.1$  & $14.3$  & $2.5$ \\
   DES16C3cje & $1.0$  & $75.3$  & $26.9$ & $1.4$  & $32.7$  & $4.5$  & $1.0$  & $24.0$  & $3.3$ \\
   \hline                                   
    & \multicolumn{3}{c}{ref.~SN: Model 6} & \multicolumn{3}{c}{ref.~SN: Model 7} & \multicolumn{3}{c}{ref.~SN: Model 8} \\    
   \hline
   OGLE073     & $26.3$  & $45.6$  & $3.8$  & $27.2$ & $32.2$  & $1.1$  & $15.1$ & $22.8$  &  $0.7$ \\
   2004ek     & $13.2$  & $26.1$  & $2.3$  & $13.7$ & $18.5$  & $0.6$  & $7.6$  & $13.0$  &  $0.4$ \\
   PTF12kso   & $5.8$   & $15.2$  & $1.8$  & $6.0$  & $10.7$  & $0.5$  & $3.3$  & $7.6$   &  $0.3$ \\
   PTF12gcx   & $5.1$   & $13.0$  & $1.2$  & $5.3$  & $9.2$   & $0.3$  & $2.9$  & $6.5$   &  $0.2$ \\
   2004em     & $10.2$  & $27.8$  & $1.1$  & $10.6$ & $19.6$  & $0.3$  & $5.9$  & $13.9$  &  $0.2$ \\
   2006V      & $2.9$   & $15.4$  & $2.0$  & $3.0$  & $10.9$  & $0.6$  & $1.7$  & $7.7$   &  $0.4$ \\
   2006au     & $9.3$   & $21.0$  & $0.8$  & $9.6$  & $14.8$  & $0.2$  & $5.3$  & $10.5$  &  $0.1$ \\
   1998A      & $14.3$  & $33.8$  & $0.6$  & $14.8$ & $23.9$  & $0.2$  & $8.2$  & $16.9$  &  $0.1$ \\
   1987A      & $1.3$   & $13.8$  & $2.2$  & $1.3$  & $9.8$   & $0.6$  & $0.7$  & $6.9$   &  $0.4$ \\
   2000cb     & $6.5$   & $16.7$  & $0.5$  & $6.7$  & $11.8$  & $0.1$  & $3.7$  & $8.3$   &  $0.1$ \\
   2005ci     & $3.2$   & $20.0$  & $0.9$  & $3.3$  & $14.1$  & $0.3$  & $1.9$  & $10.0$  &  $0.2$ \\
   2009mw     & $3.7$   & $18.7$  & $0.7$  & $3.8$  & $13.2$  & $0.2$  & $2.1$  & $9.4$   &  $0.1$ \\
   2009E      & $0.6$   & $11.9$  & $2.6$  & $0.6$  & $8.4$   & $0.7$  & $0.4$  & $5.9$   &  $0.5$ \\
   DES16C3cje & $0.6$   & $20.0$  & $3.4$  & $0.6$  & $14.1$  & $1.0$  & $0.3$  & $10.0$  &  $0.6$ \\
   \hline                                   
 \end{tabular}
 \label{tab_mod_Arnett_highNi}
\end{table*}


\begin{table*}
   \centering
   \caption{Same as Table~\ref{tab_mod_Arnett_lowNi}, but for the Popov set. The symbol $\leq$ ($\geq$) indicates that it is possible to estimate just an upper (lower) limit for the considered physical quantity (cf.~Table \ref{tab_Popov}).}
   \begin{tabular}{lccccccrrr}
   \hline\hline
   SN & \multicolumn{1}{c}{$E$} & \multicolumn{1}{c}{$M_{ej}$} & \multicolumn{1}{c}{$R$} & \multicolumn{1}{c}{$E$} & \multicolumn{1}{c}{$M_{ej}$} & \multicolumn{1}{c}{$R$} & \multicolumn{1}{c}{$\pm E$} & \multicolumn{1}{c}{$\pm M_{ej}$} & \multicolumn{1}{c}{$\pm R$}  \\
      & \multicolumn{1}{c}{[foe]} & \multicolumn{1}{c}{[\msun]} & \multicolumn{1}{c}{[$10^{12}cm$]} & \multicolumn{1}{c}{[foe]} & \multicolumn{1}{c}{[\msun]} & \multicolumn{1}{c}{[$10^{12}cm$]} & \multicolumn{1}{c}{[$\%$]} & \multicolumn{1}{c}{[$\%$]} & \multicolumn{1}{c}{[$\%$]}\\
   \hline
    & \multicolumn{3}{c}{ref.~SN: Model 1} & \multicolumn{3}{c}{ref.~SN: Model 2} & \multicolumn{3}{c}{Error percentages}\\
   \hline
   2004ek     & $0.7$        & $13.3$       & $186.3$    & $0.8$        & $12.9$       & $161.4$    & $93$  & $53$  & $43$  \\
   PTF12gcx   & $3.3$        & $28.8$       & $14.8$     & $3.7$        & $28.0$       & $12.9$     & $349$ & $230$ & $129$ \\
   2004em     & $12.9$       & $92.4$       & $18.9$     & $14.5$       & $89.9$       & $16.4$     & $370$ & $235$ & $146$ \\
   2006au     & $11.7$       & $69.9$       & $13.1$     & $13.2$       & $68.0$       & $11.3$     & $182$ & $92$  & $92$  \\
   1987A      & $6.25$       & $103.3$      & $6.4$      & $7.0$        & $100.0$      & $5.53$     & $95$  & $58$  & $66$  \\
   2000cb     & $\geq 357.6$ & $\geq 535.5$ & $\leq 0.5$ & $\geq 401.3$ & $\geq 521.0$ & $\leq 0.5$ & $98$  & $58$  & $71$  \\
   2005ci     & $\geq 140.9$ & $\geq 555.7$ & $\leq 1.2$ & $\geq 158.1$ & $\geq 540.7$ & $\leq 1.0$ & $61$  & $102$ & $80$  \\
   2009mw     & $\geq 51.2$  & $\geq 163.4$ & $\leq 2.6$ & $\geq 57.4$  & $\geq 256.3$ & $\leq 2.2$ & $110$ & $63$  & $64$  \\
   DES16C3cje & $ 2.3$       & $132.0$      & $19.6$     & $2.5$        & $128.4$      & $ 17.0$    & $635$ & $320$ & $322$ \\
   \hline                                   
 \end{tabular}
 \label{tab_mod_Popov_lowNi}
\end{table*}

\begin{table*}
   \centering
   \caption{Same as Table~\ref{tab_mod_Arnett_highNi}, but for the Popov set. See Table~\ref{tab_mod_Popov_lowNi} for the errors on the values of $E$, $M_{ej}$, and $R$.}
   \begin{tabular}{lccccccccc}
   \hline\hline
   SN & \multicolumn{1}{c}{$E$} & \multicolumn{1}{c}{$M_{ej}$} & \multicolumn{1}{c}{$R$} & \multicolumn{1}{c}{$E$} & \multicolumn{1}{c}{$M_{ej}$} & \multicolumn{1}{c}{$R$} & \multicolumn{1}{c}{$E$} & \multicolumn{1}{c}{$M_{ej}$} & \multicolumn{1}{c}{$R$} \\
      & \multicolumn{1}{c}{[foe]} & \multicolumn{1}{c}{[\msun]} & \multicolumn{1}{c}{[$10^{12}cm$]} & \multicolumn{1}{c}{[foe]} & \multicolumn{1}{c}{[\msun]} & \multicolumn{1}{c}{[$10^{12}cm$]} & \multicolumn{1}{c}{[foe]} & \multicolumn{1}{c}{[\msun]} & \multicolumn{1}{c}{[$10^{12}cm$]} \\
   \hline   
    & \multicolumn{3}{c}{ref.~SN: Model 3} & \multicolumn{3}{c}{ref.~SN: Model 4} & \multicolumn{3}{c}{ref.~SN: Model 5} \\    
   \hline
   2004ek     & $0.5$        & $10.2$       & $194.1$    & $0.4$        & $3.1$        & $74.6$     & $0.1$       & $1.2$       & $131.8$    \\
   PTF12gcx   & $2.4$        & $22.1$       & $15.5$     & $1.9$        & $6.7$        & $5.9$      & $0.4$       & $2.6$       & $10.5$     \\
   2004em     & $9.5$        & $71.1$       & $19.7$     & $7.5$        & $21.6$       & $7.6$      & $1.8$       & $8.4$       & $13.4$     \\
   2006au     & $8.6$        & $53.7$       & $13.6$     & $6.8$        & $16.4$       & $5.2$      & $1.6$       & $6.4$       & $9.3$      \\
   1987A      & $4.6$        & $79.3$       & $4.3$      & $24.2$       & $3.6$        & $1.7$      & $0.9$       & $9.5$       & $4.5$      \\
   2000cb     & $\geq 263.4$ & $\geq 411.4$ & $\leq 0.4$ & $\geq 207.1$ & $\geq 125.4$ & $\leq 0.2$ & $\geq 49.6$ & $\geq 49.1$ & $\leq 0.4$ \\
   2005ci     & $\geq 103.8$ & $\geq 427.0$ & $\leq 1.0$ & $\geq 81.6$  & $\geq 130.2$ & $\leq 0.5$ & $\geq 19.5$ & $\geq 50.9$ & $\leq 0.9$ \\
   2009mw     & $\geq 37.7$  & $\geq 202.4$ & $\leq 2.7$ & $\geq 29.6$  & $\geq 61.7$  & $\leq 1$   & $\geq 7.1$  & $\geq 24.1$ & $\leq 1.8$ \\
   DES16C3cje & $1.7$        & $101.4$      & $20.4$     & $1.3$        & $30.9$       & $7.8$      & $0.3$       & $12.1$      & $13.9$     \\
   \hline                                   
    & \multicolumn{3}{c}{ref.~SN: Model 6} & \multicolumn{3}{c}{ref.~SN: Model 7} & \multicolumn{3}{c}{ref.~SN: Model 8} \\    
   \hline
   2004ek     & $4.0\mbox{E-2}$& $0.8$        & $224.0$     & $0.02$     & $0.4$       & $187.9$    &  $1.0\mbox{E-2}$& $0.1$      & $142.5$    \\
   PTF12gcx   & $0.1$          & $1.5$        & $17.9$      & $0.1$      & $0.8$       & $15.0$     &  $2.0\mbox{E-2}$& $0.3$      & $26.3$     \\
   2004em     & $0.5$          & $4.7$        & $22.8$      & $0.4$      & $2.5$       & $19.1$     &  $7.0\mbox{E-2}$& $1.0$      & $48.9$     \\
   2006au     & $0.5$          & $3.6$        & $15.7$      & $0.3$      & $1.9$       & $13.2$     &  $7.0\mbox{E-2}$& $0.8$      & $21.3$     \\
   1987A      & $0.2$          & $5.3$        & $5.0$       & $0.2$      & $2.8$       & $4.2 $     &  $4.0\mbox{E-2}$& $1.1$      & $11.3$     \\
   2000cb     & $\geq 14.7$    & $\geq 27.3$  & $\leq 0.5$  & $\geq 9.7$ & $\geq 14.7$ & $\leq 0.6$ &  $\geq 2.0$     & $\geq 5.8$ & $\leq 1.0$ \\
   2005ci     & $\geq 5.8$     & $\geq 28.4$  & $\leq 1.5$  & $\geq 3.8$ & $\geq 15.3$ & $\leq 0.9$ &  $\geq 0.8$     & $\geq 6.0$ & $\leq 1.7$ \\
   2009mw     & $\geq 2.1$     & $\geq 13.5$  & $\leq 0.31$ & $\geq 1.4$ & $\geq 7.24$ & $\leq 0.5$ &  $\geq 0.3$     & $\geq 2.8$ & $\leq 4.6$ \\
   DES16C3cje & $0.1$          & $6.7$        & $23.6$      & $0.1$      & $3.6$       & $19.8$     &  $1.0\mbox{E-2}$& $1.4$      & $112.8$    \\  
   \hline                                   
 \end{tabular}
 \label{tab_mod_Popov_highNi}
\end{table*}


\begin{table*}
   \centering
   \caption{Same as Table~\ref{tab_mod_Arnett_lowNi}, but for the ``pure photometric'' Popov set. The symbol $\leq$ ($\geq$) indicates that it is possible to estimate just an upper (lower) limit for the considered physical quantity (cf.~Table \ref{tab_Popov}).}
   \begin{tabular}{lcccccccrrrr}
   \hline\hline
   
   SN & \multicolumn{1}{c}{$E$} & \multicolumn{1}{c}{$M_{ej}$} & \multicolumn{1}{c}{$R$} & \multicolumn{1}{c}{$E$} & \multicolumn{1}{c}{$M_{ej}$} & \multicolumn{1}{c}{$R$} & & \multicolumn{1}{c}{$\pm E$} & \multicolumn{1}{c}{$\pm M_{ej}$} & \multicolumn{1}{c}{$\pm R$}  \\
      & \multicolumn{1}{c}{[foe]} & \multicolumn{1}{c}{[\msun]} & \multicolumn{1}{c}{[$10^{12}cm$]} & \multicolumn{1}{c}{[foe]} & \multicolumn{1}{c}{[\msun]} & \multicolumn{1}{c}{[$10^{12}cm$]} & & \multicolumn{1}{c}{[$\%$]} & \multicolumn{1}{c}{[$\%$]} & \multicolumn{1}{c}{[$\%$]}\\    
     
   \hline
   & \multicolumn{3}{c}{ref.~SN: Model 1} & \multicolumn{3}{c}{ref.~SN: Model 2} & & \multicolumn{3}{c}{Error percentages}\\
   \hline
   2004ek     & $12.4$      & $54.8
 $      & $45.1
$     & $12.6$      & $50.7
 $      & $41.1
$     & & $ 17$ & $12
$ & $ 19
$ \\
   PTF12gcx   & $45.7$      & $107.6
$      & $ 4.0
$     & $46.4$      & $99.6
 $      & $ 3.6
$     & & $104$ & $65
$ & $ 99
$ \\
   2004em     & $5.9
 $      & $62.3
 $      & $28.1$     & $6.0
 $      & $57.7 $      & $25.6
$     & & $134$ & $77
$ & $105
$ \\
   2006au     & $4.6
 $      & $43.6
 $      & $21.0$     & $4.6
 $      & $40.4
 $      & $19.1$     & & $ 81$ & $42
$ & $ 45
$ \\
   1987A      & $40.3$      & $262.4
$      & $ 2.5
$     & $41.0$      & $242.9$      & $ 2.3$
     & & $ 89
$ & $52
$ & $ 67
$ \\
   2000cb     & $\geq 36.9$ & $\geq 172.0
$ & $\leq 1.7
$ & $\geq 37.5
$ & $\geq 159.2$ & $\leq 1.5
$ & & $148$ & $79
$ & $ 91
$ \\
   2005ci     & $\geq 17.8$ & $\geq 197.3
$ & $\leq 3.4
$ & $\geq 18.1$ & $\geq 182.7$ & $\leq 3.1$
 & & $132
$ & $74
$ & $ 90
$ \\
   2009mw     & $\geq 5.6
 $ & $\geq 86.9
 $ & $\leq 7.8$ & $\geq 5.7
 $ & $\geq 80.5
 $ & $\leq 7.1
$ & & $ 64$ & $38$ & $ 51
$ \\
   DES16C3cje & $1.9 $      & $119.8$      & $21.6$     & $1.9 $      & $110.9$      & $19.7$     & & $180$ & $95$ & $107$ \\
   \hline                                   
 \end{tabular}
 \label{tab_mod_PurePhot_Popov_lowNi}
\end{table*}

\begin{table*}
   \centering
   \caption{Same as Table~\ref{tab_mod_Arnett_highNi}, but for the ``pure photometric'' Popov set. See Table~\ref{tab_mod_PurePhot_Popov_lowNi} for the errors on the values of $E$, $M_{ej}$, and $R$.}
   \begin{tabular}{lccccccccc}
   \hline\hline
   SN & \multicolumn{1}{c}{$E$} & \multicolumn{1}{c}{$M_{ej}$} & \multicolumn{1}{c}{$R$} & \multicolumn{1}{c}{$E$} & \multicolumn{1}{c}{$M_{ej}$} & \multicolumn{1}{c}{$R$} & \multicolumn{1}{c}{$E$} & \multicolumn{1}{c}{$M_{ej}$} & \multicolumn{1}{c}{$R$} \\
      & \multicolumn{1}{c}{[foe]} & \multicolumn{1}{c}{[\msun]} & \multicolumn{1}{c}{[$10^{12}cm$]} & \multicolumn{1}{c}{[foe]} & \multicolumn{1}{c}{[\msun]} & \multicolumn{1}{c}{[$10^{12}cm$]} & \multicolumn{1}{c}{[foe]} & \multicolumn{1}{c}{[\msun]} & \multicolumn{1}{c}{[$10^{12}cm$]} \\
   \hline   
    & \multicolumn{3}{c}{ref.~SN: Model 3} & \multicolumn{3}{c}{ref.~SN: Model 4} & \multicolumn{3}{c}{ref.~SN: Model 5} \\    
   \hline
   2004ek     & $6.9 $      & $36.5 $      & $54.2$      & $1.2$      & $5.3 $      & $43.8$      & $0.3$      & $1.9$      & $82.6$      \\
   PTF12gcx   & $25.3$      & $71.7 $      & $4.8 $      & $4.5$      & $10.4$      & $3.9 $      & $0.9$      & $3.8$      & $7.3
 $      \\
   2004em     & $3.3 $      & $41.5 $      & $33.7$      & $0.6$      & $6.0 $      & $27.2$      & $0.1$      & $2.2$      & $51.4$      \\
   2006au     & $2.5 $      & $29.0 $      & $25.2$      & $0.4$      & $4.2 $      & $20.4$      & $0.1$      & $1.5$      & $38.4$      \\
   1987A      & $22.4$      & $174.8$      & $3.0 $      & $4.0$      & $25.4$      & $2.4 $      & $0.8$      & $9.3$      & $4.6
 $      \\
   2000cb     & $\geq 20.4$ & $\geq 114.6$ & $\leq 2.0 $ & $\geq 3.6$ & $\geq 16.6$ & $\leq 1.6 $ & $\geq 0.8$ & $\geq 6.1$ & $\leq 3.1
 $ \\
   2005ci     & $\geq 9.8 $ & $\geq 131.5$ & $\leq 4.1 $ & $\geq 1.8$ & $\geq 19.1$ & $\leq 3.3 $ & $\geq 0.4$ & $\geq 7.0$ & $\leq 6.2
 $ \\
   2009mw     & $\geq 3.1 $ & $\geq 57.9 $ & $\leq 9.4 $ & $\geq 0.5$ & $\geq 8.4 $ & $\leq 7.6 $ & $\geq 0.1$ & $\geq 3.1$ & $\leq 14.3$ \\
   DES16C3cje & $1.0 $      & $79.9 $      & $25.9$      & $0.2$      & $11.6$      & $21.0$      & $0.0$      & $4.2$      & $39.6$      \\
   \hline
   & \multicolumn{3}{c}{ref.~SN: Model 6} & \multicolumn{3}{c}{ref.~SN: Model 7} & \multicolumn{3}{c}{ref.~SN: Model 8} \\
   \hline
   2004ek     & $9.0\mbox{E-2}$     & $1.2$      & $129.2$      & $3.2\mbox{E-3}$      & $0.1$      & $466.1$      & $3.0\mbox{E-4}$      & $4.0\mbox{E-2}$     & $1224.6$     \\
   PTF12gcx   & $3.3\mbox{E-2}$     & $2.3$      & $11.4 $      & $1.2\mbox{E-2}$      & $0.3$      & $41.0 $      & $1.1\mbox{E-3}$      & $8.0\mbox{E-2}$     & $107.8
 $     \\
   2004em     & $4.0\mbox{E-2}$     & $1.3$      & $80.3 $      & $1.5\mbox{E-3}$      & $0.2$      & $289.8$      & $1.4\mbox{E-4}$      & $4.0\mbox{E-2}$     & $761.4
 $     \\
   2006au     & $3.0\mbox{E-2}$     & $0.9$      & $60.0 $      & $1.2\mbox{E-3}$      & $0.1$      & $216.7$      & $1.1\mbox{E-4}$      & $3.0\mbox{E-2}$     & $569.3
 $     \\
   1987A      & $2.9\mbox{E-2}$     & $5.6$      & $7.2  $      & $1.0\mbox{E-2}$      & $0.7$      & $25.9 $      & $9.6\mbox{E-4}$      & $0.2$               & $68.1
  $     \\
   2000cb     & $\geq 0.3$          & $\geq 3.7$ & $\leq 4.8  $ & $\geq 9.5\mbox{E-3}$ & $\geq 0.5$ & $\leq 17.4 $ & $\geq 8.8\mbox{E-4}$ & $\geq 0.1$          & $\leq 45.7
 $ \\
   2005ci     & $\geq 0.1$          & $\geq 4.2$ & $\leq 9.7  $ & $\geq 4.6\mbox{E-3}$ & $\geq 0.5$ & $\leq 35.2 $ & $\geq 4.2\mbox{E-4}$ & $\geq 0.1$          & $\leq 92.4
 $ \\
   2009mw     & $\geq 4.0\mbox{E-2}$& $\geq 1.9$ & $\leq 22.3 $ & $\geq 1.4\mbox{E-3}$ & $\geq 0.2$ & $\leq 80.5 $ & $\geq 1.3\mbox{E-4}$ & $\geq 6.0\mbox{E-2}$& $\leq 211.5$ \\
   DES16C3cje & $1.0\mbox{E-2}$     & $2.6$      & $61.8 $      & $4.8\mbox{E-4}$      & $0.3$      & $223.1$      & $4.5\mbox{E-5}$      & $0.08$              & $586.2 $     \\
   \hline                                   
 \end{tabular}
 \label{tab_mod_PurePhot_Popov_highNi}
\end{table*}


\begin{table*}
   \centering
   \caption{Values of $E$, $M_{ej}$ and $R$ derived from the Arnett set for the grid of radiation-hydrodynamical models (see Table \ref{tab_parameters_modelling}). The models parameters of the reference SN are put between square brackets.}
   \begin{tabular}{lcccccccccccc}
   \hline\hline
   SN & \multicolumn{1}{c}{$E$} & \multicolumn{1}{c}{$M_{ej}$} & \multicolumn{1}{c}{$R$} & \multicolumn{1}{c}{$E$} & \multicolumn{1}{c}{$M_{ej}$} & \multicolumn{1}{c}{$R$} & \multicolumn{1}{c}{$E$} & \multicolumn{1}{c}{$M_{ej}$} & \multicolumn{1}{c}{$R$}  & \multicolumn{1}{c}{$E$} & \multicolumn{1}{c}{$M_{ej}$} & \multicolumn{1}{c}{$R$} \\
      & \multicolumn{1}{c}{[foe]} & \multicolumn{1}{c}{[\msun]} & \multicolumn{1}{c}{[$10^{12}cm$]} & \multicolumn{1}{c}{[foe]} & \multicolumn{1}{c}{[\msun]} & \multicolumn{1}{c}{[$10^{12}cm$]} & \multicolumn{1}{c}{[foe]} & \multicolumn{1}{c}{[\msun]} & \multicolumn{1}{c}{[$10^{12}cm$]} & \multicolumn{1}{c}{[foe]} & \multicolumn{1}{c}{[\msun]} & \multicolumn{1}{c}{[$10^{12}cm$]}\\
   \hline   
    & \multicolumn{3}{c}{ref.~SN: Model 1} & \multicolumn{3}{c}{ref.~SN: Model 2} & \multicolumn{3}{c}{ref.~SN: Model 3}  & \multicolumn{3}{c}{ref.~SN: Model 4} \\    
   \hline
    Model 1 & $[1.0$ & $16.0$  & $3.0]$   &  $1.1$  & $15.2$  & $2.6$   & $0.9$  & $14.2$  & $2.4$   &  $1.3$  & $6.1$  & $0.4 $ \\ 
    Model 2 & $0.9$  & $16.8$  & $3.4$    &  $[1.0$ & $16.0$  & $3.0]$  & $0.9$  & $14.9$  & $2.8$   &  $1.2$  & $6.5$  & $0.5 $ \\ 
    Model 3 & $1.1$  & $18.1$  & $3.7$    &  $1.2$  & $17.2$  & $3.3$   & $[1.0$ & $16.0$  & $3.0]$  &  $1.4$  & $6.9$  & $0.5 $ \\ 
    Model 4 & $0.8$  & $41.7$  & $22.6$   &  $0.8$  & $39.7$  & $19.6$  & $0.7$  & $36.9$  & $18.1$  &  $[1.0$ & $16.0$ & $3.0]$ \\ 
    Model 5 & $1.1$  & $56.7$  & $30.6$   &  $1.2$  & $53.9$  & $26.6$  & $1.0$  & $50.2$  & $24.6$  &  $1.5$  & $21.8$ & $4.1$  \\
    Model 6 & $1.9$  & $68.1$  & $29.4$   &  $2.1$  & $64.8$  & $25.6$  & $1.8$  & $60.3$  & $23.6$  &  $2.5$  & $26.2$ & $3.9$  \\
    Model 7 & $1.8$  & $96.4$  & $104.6$  &  $2.0$  & $91.7$  & $91.1$  & $1.7$  & $85.3$  & $84.0$  &  $2.4$  & $37.0$ & $13.9$ \\
    Model 8 & $3.3$  & $136.4$ & $160.3$  &  $3.6$  & $129.8$ & $139.6$ & $3.1$  & $120.7$ & $128.7$ &  $4.4$  & $52.4$ & $21.3$ \\
   \hline   
    & \multicolumn{3}{c}{ref.~SN: Model 5} & \multicolumn{3}{c}{ref.~SN: Model 6} & \multicolumn{3}{c}{ref.~SN: Model 7}  & \multicolumn{3}{c}{ref.~SN: Model 8} \\    
   \hline
    Model 1 & $0.9$  & $4.5$  & $0.3$  & $0.5$  & $3.8$  & $0.3$  & $0.5$  & $2.7$  & $0.1$  & $ 0.3$ & $1.9$  & $0.1$  \\
    Model 2 & $0.8$  & $4.7$  & $0.3$  & $0.5$  & $3.9$  & $0.4$  & $0.5$  & $2.8$  & $0.1$  & $ 0.3$ & $2.0$  & $0.1$  \\
    Model 3 & $1.0$  & $5.1$  & $0.4$  & $0.6$  & $4.2$  & $0.4$  & $0.6$  & $3.0$  & $0.1$  & $ 0.3$ & $2.1$  & $0.1$  \\
    Model 4 & $0.7$  & $11.8$ & $2.2$  & $0.4$  & $9.8$  & $2.3$  & $0.4$  & $6.9$  & $0.6$  & $ 0.2$ & $4.9$  & $0.4$  \\
    Model 5 & $[1.0$ & $16.0$ & $3.0]$ & $0.6$  & $13.3$ & $3.1$  & $0.6$  & $9.4$  & $0.9$  & $0.3$  & $6.6$  & $0.6$  \\
    Model 6 & $1.7$  & $19.2$ & $2.9$  & $[1.0$ & $16.0$ & $3.0]$ & $1.0$  & $11.3$ & $0.8$  & $0.6$  & $8.0$  & $0.6$  \\
    Model 7 & $1.7$  & $27.2$ & $10.3$ & $1.0$  & $22.6$ & $10.7$ & $[1.0$ & $16.0$ & $3.0]$ & $0.6$  & $11.3$ & $2.0$  \\
    Model 8 & $3.0$  & $38.5$ & $15.7$ & $1.7$  & $32.0$ & $16.3$ & $1.8$  & $22.6$ & $4.6$  & $[1.0$ & $16.0$ & $3.0]$ \\
   \hline     
   \end{tabular}
\label{tab_grid_mod_Arnett}
\end{table*}

\begin{table*}
   \centering
   \caption{Same as Table~\ref{tab_grid_mod_Arnett}, but for the Popov set.}
   \begin{tabular}{lcccccccccccc}
   \hline\hline
   SN & \multicolumn{1}{c}{$E$} & \multicolumn{1}{c}{$M_{ej}$} & \multicolumn{1}{c}{$R$} & \multicolumn{1}{c}{$E$} & \multicolumn{1}{c}{$M_{ej}$} & \multicolumn{1}{c}{$R$} & \multicolumn{1}{c}{$E$} & \multicolumn{1}{c}{$M_{ej}$} & \multicolumn{1}{c}{$R$}  & \multicolumn{1}{c}{$E$} & \multicolumn{1}{c}{$M_{ej}$} & \multicolumn{1}{c}{$R$} \\
      & \multicolumn{1}{c}{[foe]} & \multicolumn{1}{c}{[\msun]} & \multicolumn{1}{c}{[$10^{12}cm$]} & \multicolumn{1}{c}{[foe]} & \multicolumn{1}{c}{[\msun]} & \multicolumn{1}{c}{[$10^{12}cm$]} & \multicolumn{1}{c}{[foe]} & \multicolumn{1}{c}{[\msun]} & \multicolumn{1}{c}{[$10^{12}cm$]} & \multicolumn{1}{c}{[foe]} & \multicolumn{1}{c}{[\msun]} & \multicolumn{1}{c}{[$10^{12}cm$]}\\
   \hline   
    & \multicolumn{3}{c}{ref.~SN: Model 1} & \multicolumn{3}{c}{ref.~SN: Model 2} & \multicolumn{3}{c}{ref.~SN: Model 3}  & \multicolumn{3}{c}{ref.~SN: Model 4} \\    
   \hline
    Model 1 & $[1.0$  & $16.0$   & $3.0]$ & $1.1$   & $15.6$   & $2.6$  & $0.7$   & $12.3$   & $3.1$  & $0.6$   & $3.7$   & $1.2$  \\ 
    Model 2 & $0.9$   & $16.4$   & $3.5$  & $[1.0$  & $16.0$   & $3.0]$ & $0.7$   & $12.6$   & $3.6$  & $0.5$   & $3.9$   & $1.4$  \\ 
    Model 3 & $1.4$   & $20.8$   & $2.9$  & $1.5$   & $20.3$   & $2.5$  & $[1.0$  & $16.0$   & $3.0]$ & $0.8$   & $4.9$   & $1.2$  \\ 
    Model 4 & $1.7$   & $68.3$   & $7.5$  & $1.9$   & $66.5$   & $6.5$  & $1.3$   & $52.5$   & $7.8$  & $[1.0$  & $16.0$  & $3.0]$ \\ 
    Model 5 & $7.2$   & $174.6$  & $4.2$  & $8.1$   & $169.9$  & $3.7$  & $5.3$   & $134.1$  & $4.4$  & $4.2$   & $40.9$  & $1.7$  \\
    Model 6 & $24.3$  & $313.4$  & $2.5$  & $27.2$  & $304.9$  & $2.2$  & $17.9$  & $240.8$  & $2.6$  & $14.1$  & $73.4$  & $1.0$  \\
    Model 7 & $36.9$  & $582.1$  & $3.0$  & $41.4$  & $566.4$  & $2.6$  & $27.1$  & $447.2$  & $3.1$  & $21.4$  & $136.4$ & $1.2$  \\
    Model 8 & $178.3$ & $1488.1$ & $1.7$  & $200.1$ & $1447.9$ & $1.5$  & $131.3$ & $1143.2$ & $1.8$  & $103.3$ & $348.6$ & $0.7$  \\
   \hline   
    & \multicolumn{3}{c}{ref.~SN: Model 5} & \multicolumn{3}{c}{ref.~SN: Model 6} & \multicolumn{3}{c}{ref.~SN: Model 7}  & \multicolumn{3}{c}{ref.~SN: Model 8} \\    
   \hline
    Model 1 & $0.1$  & $1.5$   & $2.1$  & $0.04$ & $0.8$  & $3.6$  & $0.03$ & $0.4$  & $3.0$  & $0.006$ & $0.2$  & $5.3$  \\ 
    Model 2 & $0.1$  & $1.5$   & $2.4$  & $0.04$ & $0.8$  & $4.2$  & $0.02$ & $0.5$  & $3.5$  & $0.005$ & $0.2$  & $6.1$  \\ 
    Model 3 & $0.2$  & $1.9$   & $2.0$  & $0.06$ & $1.1$  & $3.5$  & $0.04$ & $0.6$  & $2.9$  & $0.01$  & $0.2$  & $5.1$  \\ 
    Model 4 & $0.2$  & $6.3$   & $5.3$  & $0.07$ & $3.5$  & $9.0$  & $0.05$ & $1.9$  & $7.6$  & $0.01$  & $0.7$  & $13.3$ \\ 
    Model 5 & $[1.0$ & $16.0$  & $3.0]$ & $0.3$  & $8.9$  & $5.1$  & $0.2$  & $4.8$  & $4.3$  & $0.04$  & $1.9$  & $7.5$  \\
    Model 6 & $3.4$  & $28.7$  & $1.8$  & $[1.0$ & $16.0$ & $3.0]$ & $0.7$  & $8.6$  & $2.5$  & $0.14$  & $3.4$  & $4.4$  \\
    Model 7 & $5.1$  & $53.3$  & $2.1$  & $1.5$  & $29.7$ & $3.6$  & $[1.0$ & $16.0$ & $3.0]$ & $0.21$  & $6.3$  & $5.3$  \\
    Model 8 & $24.7$ & $136.4$ & $1.2$  & $7.3$  & $76.0$ & $2.0$  & $4.8$  & $40.9$ & $1.7$  & $[1.0$  & $16.0$ & $3.0]$ \\
   \hline     
   \end{tabular}
\label{tab_grid_mod_Popov}
\end{table*}

\begin{table*}
   \centering
   \caption{Same as Table~\ref{tab_grid_mod_Arnett}, but for the ``pure photometric'' Popov set.}
   \begin{tabular}{lcccccccccccc}
   \hline\hline
   SN & \multicolumn{1}{c}{$E$} & \multicolumn{1}{c}{$M_{ej}$} & \multicolumn{1}{c}{$R$} & \multicolumn{1}{c}{$E$} & \multicolumn{1}{c}{$M_{ej}$} & \multicolumn{1}{c}{$R$} & \multicolumn{1}{c}{$E$} & \multicolumn{1}{c}{$M_{ej}$} & \multicolumn{1}{c}{$R$}  & \multicolumn{1}{c}{$E$} & \multicolumn{1}{c}{$M_{ej}$} & \multicolumn{1}{c}{$R$} \\
      & \multicolumn{1}{c}{[foe]} & \multicolumn{1}{c}{[\msun]} & \multicolumn{1}{c}{[$10^{12}cm$]} & \multicolumn{1}{c}{[foe]} & \multicolumn{1}{c}{[\msun]} & \multicolumn{1}{c}{[$10^{12}cm$]} & \multicolumn{1}{c}{[foe]} & \multicolumn{1}{c}{[\msun]} & \multicolumn{1}{c}{[$10^{12}cm$]} & \multicolumn{1}{c}{[foe]} & \multicolumn{1}{c}{[\msun]} & \multicolumn{1}{c}{[$10^{12}cm$]}\\
   \hline   
    & \multicolumn{3}{c}{ref.~SN: Model 1} & \multicolumn{3}{c}{ref.~SN: Model 2} & \multicolumn{3}{c}{ref.~SN: Model 3}  & \multicolumn{3}{c}{ref.~SN: Model 4} \\    
   \hline
   Model 1 & $[1.0   $ & $16.0   $ & $3.0]$& $1.0    $ & $14.8   $ & $2.7$ & $0.6    $ & $10.7   $ & $3.6$ & $0.1   $ & $1.5   $ & $2.9
$  \\
   Model 2 & $1.0    $ & $17.3   $ & $3.3$ & $[1.0   $ & $16.0   $ & $3.0]$& $0.5    $ & $11.5   $ & $4.0$ & $0.1   $ & $1.7   $ & $3.2
$  \\
   Model 3 & $1.8    $ & $24.0   $ & $2.5$ & $1.8    $ & $22.2   $ & $2.3$ & $[1.0   $ & $16.0   $ & $3.0]$& $0.2   $ & $2.3   $ & $2.4
$  \\
   Model 4 & $10.1   $ & $165.5  $ & $3.1$ & $10.3   $ & $153.3  $ & $2.8$ & $5.6    $ & $110.3  $ & $3.7$ & $[1.0  $ & $16.0  $ & $3.0
]$ \\
   Model 5 & $48.3   $ & $451.8  $ & $1.6$ & $49.1   $ & $418.3  $ & $1.5$ & $26.8   $ & $301.1  $ & $2.0$ & $4.8   $ & $43.7  $ & $1.6
$  \\
   Model 6 & $137.5  $ & $745.8  $ & $1.0$ & $139.6  $ & $690.4  $ & $1.0$ & $76.2   $ & $496.9  $ & $1.3$ & $13.6  $ & $72.1  $ & $1.0
$  \\
   Model 7 & $3865.4 $ & $5960.5 $ & $0.3$ & $3926.6 $ & $5518.2 $ & $0.3$ & $2141.4 $ & $3971.6 $ & $0.3$ & $381.1 $ & $576.1 $ & $0.3
$  \\
   Model 8 & $41807.5$ & $22783.9$ & $0.1$ & $42469.4$ & $21093.1$ & $0.1$ & $23161.2$ & $15181.4$ & $0.1$ & $4122.3$ & $2202.1$ & $0.1$  \\
   \hline   
    & \multicolumn{3}{c}{ref.~SN: Model 5} & \multicolumn{3}{c}{ref.~SN: Model 6} & \multicolumn{3}{c}{ref.~SN: Model 7}  & \multicolumn{3}{c}{ref.~SN: Model 8} \\    
   \hline
   Model 1 & $2.0\mbox{E-2}$& $0.6  $ & $5.5$ & $0.01  $ & $0.3  $ & $8.6$ & $2.6\mbox{E-4}$& $4.0\mbox{E-2}$& $31.0$ & $2.4\mbox{E-5}$& $1.0\mbox{E-2}$& $81.4$ \\
   Model 2 & $2.0\mbox{E-2}$& $0.6  $ & $6.0$ & $0.01  $ & $0.4  $ & $9.4$ & $2.5\mbox{E-4}$& $5.0\mbox{E-2}$& $34.0$ & $2.4\mbox{E-5}$& $1.0\mbox{E-2}$& $89.4$
 \\
   Model 3 & $4.0\mbox{E-2}$& $0.9  $ & $4.6$ & $0.01  $ & $0.5  $ & $7.1$ & $4.7\mbox{E-4}$& $6.0\mbox{E-2}$& $25.8$ & $4.3\mbox{E-5}$& $2.0\mbox{E-2}$& $67.8$
 \\
   Model 4 & $0.2  $        & $5.9  $ & $5.7$ & $0.07  $ & $3.6  $ & $8.9$ & $2.6\mbox{E-3}$& $0.4 $         & $31.9$ & $2.4\mbox{E-4}$& $0.1 $         & $83.9$
 \\
   Model 5 & $[1.0 $        & $16.0 $ & $3.0]$& $0.35  $ & $9.7  $ & $4.7$ & $1.2\mbox{E-2}$& $1.2 $         & $16.9$ & $1.2\mbox{E-3}$& $0.3 $         & $44.5$
 \\
   Model 6 & $2.8  $        & $26.4 $ & $1.9$ & $[1.0  $ & $16.0 $ & $3.0]$& $3.6\mbox{E-2}$& $2.0 $         & $10.8$ & $3.3\mbox{E-3}$& $0.5 $         & $28.4$
 \\
   Model 7 & $80.0 $        & $211.1$ & $0.5$ & $28.12 $ & $127.9$ & $0.8$ & $[1.0$         & $16.0$         & $3.0]$ & $9.2\mbox{E-2}$& $4.2 $         & $7.9
$  \\
   Model 8 & $865.0$        & $806.8$ & $0.2$ & $304.13$ & $488.8$ & $0.3$ & $10.8$         & $61.2$         & $1.1 $ & $[1.0$         & $16.0$         & $3.0]$  \\
    \hline     
   \end{tabular}
\label{tab_grid_mod_PurePhot_Popov}
\end{table*}

\label{lastpage}
\end{document}